\documentclass{IEEEtran}
\usepackage{cite}
\usepackage{amsmath,amssymb,amsfonts}
\usepackage{graphicx}
\usepackage{tikz,pgfplots}
\usepackage{multirow}
\usepackage{subcaption}
\usepackage{textcomp,nicefrac}
\usepackage{lineno}
\usepackage{url}
\newcommand{\niss}[1]{_{\textnormal{#1}}} %NonItalicSubScript
\def\BibTeX{{\rm B\kern-.05em{\sc i\kern-.025em b}\kern-.08em
T\kern-.1667em\lower.7ex\hbox{E}\kern-.125emX}}
\begin{document}
% \linenumbers
\bstctlcite{IEEEexample:BSTcontrol}
\title{Operation of a negative ion gas time projection chamber without electronegative fill gases}
% \author{\IEEEauthorblockN{
% Lachlan J. McKie\IEEEauthorrefmark{1}\IEEEauthorrefmark{3},
% Lindsey J. Bignell\IEEEauthorrefmark{2}\IEEEauthorrefmark{3},
% Nicholas Adams\IEEEauthorrefmark{3},
% Victoria U. Bashu\IEEEauthorrefmark{3},
% Ferdos Dastgiri\IEEEauthorrefmark{3},
% Gregory J. Lane\IEEEauthorrefmark{3}}
                            
% \IEEEauthorblockA{\IEEEauthorrefmark{1}Corresponding Author. E-mail: mckiel@ansto.gov.au}

% \IEEEauthorblockA{\IEEEauthorrefmark{2}Corresponding Author. E-mail: lindsey.bignell@anu.edu.au}

% \IEEEauthorblockA{\IEEEauthorrefmark{3} Centre of Excellence for Dark Matter Particle Physics, Research School of Physics, The Australian National University, Acton ACT 2602, Australia}

\author{Lachlan J. McKie, Lindsey J. Bignell, Nicholas Adams, Victoria U. Bashu, Ferdos Dastgiri and Gregory J. Lane
\thanks{This work was supported by the Australian Research
Council under grant number CE200100008 and the Australian Government through Defence under grant 12138. Support for the ANU Heavy Ion Accelerator Facility operations through the Australian National Collaborative Research Infrastructure Strategy (NCRIS) program. LJM additionally acknowledges the support of the Australian Government Research Training Program. \textit{Corresponding authors: L. J. McKie (mckiel@ansto.gov.au) and L. J. Bignell (lindsey.bignell@anu.edu.au).}}
\thanks{L. J. McKie, L. J. Bignell, N. Adams, V. U. Bashu, F. Dastgiri, and G. J. Lane are with the Centre of Excellence for Dark Matter Particle Physics, Research School of Physics, The Australian National University, Acton ACT 2601 Australia.}
\thanks{L. J. McKie is now with the Centre for Accelerator Science at the Australian Nuclear Science and Technology Organisation~(ANSTO), Sydney, NSW 2232 Australia. F. Dastgiri is now with University College London, Department of Physics and Astronomy, WC1E 6BT, United Kingdom.}}

\maketitle

\begin{abstract}
The high fidelity reconstruction of particle tracks in micropatterned gaseous time projection chambers renders this technology ideal for future rare-event searches, including direction-sensitive dark matter experiments. Large drift distances are typically required for such experiments, so that the overall spatial resolution is limited by diffusion. Negative ion drift exhibits lower diffusion than electron drift and is thus an attractive option for realising a large-scale detector. The use of electronegative gases to create negative ions introduces technical challenges, most notably a reduction in gain when compared to conventional gas mixtures. In this study, we demonstrate a new method for negative ion generation via dissociative electron attachment using the conventional molecular fill gas CF$_4$. Our optical measurements of negative ion drift indicate electron attachment lengths of $<$~1~mm and comparable gain to electron avalanches. The individual negative ion avalanches were also time-resolved, allowing the number of ions reaching the readout to be counted. We measure an improved energy resolution by single ion counting, relative to an integrated electron avalanche signal measured under identical gain conditions. 
\end{abstract}

\begin{IEEEkeywords}
CF$_4$, CYGNUS, Dark matter, Direction-sensitive detectors, Dissociative electron attachment, Gaseous time projection chamber, Negative ion drift 
\end{IEEEkeywords}

\section{Introduction}
\label{sec:introduction}
\label{sec:intro}

\IEEEPARstart{G}{as} Time Projection Chambers~(TPCs) offer a promising means to achieve direction-sensitive detectors for rare event physics experiments, with possible applications in the search for Weakly Interacting Massive Particle~(WIMP) Dark Matter~\cite{Vahsen2020, vahsen2021directional}, solar neutrino physics~\cite{lisotti2024cygnus}, and other applications~\cite{OHare2022Recoil}. Electron diffusion limits the achievable spatial resolution in large-scale gaseous time projection chambers, posing a challenge for applications which require fine track features to be preserved at the readout. Negative Ion Drift~(NID) offers a promising alternative to minimise diffusion. NID involves the capture of ionisation electrons onto a gas species to form negative ions, which drift to the readout with low diffusion, at the thermal limit under suitable conditions~\cite{Martoff2000, Martoff2005}. Studies of NID have to date required the introduction of electronegative fill gases, with the effect reported for CS$_2$~\cite{Alner2005a, Battat2015, Snowden-ifft2000}, O$_2$~\cite{Sorensen2012}, CH$_3$NO$_2$~\cite{Martoff2009}, and SF$_6$~\cite{Phan, Amaro2024}. However, the use of such gases introduces technical challenges within detectors; gain tends to be lower and gases such as CS$_2$ and CH$_3$NO$_2$ pose significant safety challenges for a large underground experiment owing to their flammability and toxicity. While multi-mesh thick Gas Electron Multipliers (GEMs) offer a promising means to improve the gain characteristics of NID detectors~\cite{Eldridge2023,McLean2024_, McLean_2026_JINST_21_P07047}, the pitches of these gain structures are $>1$~mm, which limits their applicability as a means to preserve track features below this size. A benign, high-gain, NID-capable gas mixture is therefore a worthy goal for future experimental projects. 

The formation of gas-phase negative ions through electron attachment has been well-studied~\cite{Massey1976, McDaniel1964}, with various attachment processes identified. The primary attachment mechanism reported in gaseous TPCs so far is attributed to \textit{three-body attachment}, where an electronegative gas species captures a primary electron in an exothermic process to create an excited Temporary Negative Ion~(TNI), which transfers its excess energy to a third body via collisional stabilisation. This process typically requires low electron energies, approximately~($\varepsilon<1$~eV). Three-body attachment is expected to be the dominant attachment process at low reduced fields. Another mechanism, \textit{Dissociative Electron Attachment}~(DEA), has been proposed to explain the experimental observation of minority negative ion carriers in SF$_6$~\cite{Phan}. DEA involves the fragmentation of a molecular gas species into a neutral and negative ion radical. Minority carriers are transported with a different drift velocity, allowing for detector fiducialisation~\cite{Battat2015, Battat2016} . Unlike three-body attachment, DEA typically occurs at higher electron energies, sufficient to sever the intramolecular bond and liberate fragments of the parent molecule~\cite{Massey1976}. In the case of SF$_6$, the DEA threshold is approximately 1~eV and shares an overlapping cross section with three-body attachment~\cite{Christophorou2001}. More typically, these processes are energetically separated, with a larger electron energy required for DEA, resulting in an operational threshold before DEA is observable. The DEA threshold often remains lower than the ionization energy of the gas and is not exclusive to electronegative gases, so may be achieved by conventional molecular TPC gases under suitable detector conditions. In this work, CF$_4$, a common TPC fill gas, was selected for investigation as a NID gas using DEA. CF$_4$ is widely used in TPCs due to its excellent scintillation properties and high drift velocity~\cite{Pansky1995} and previous experimental observations with CF$_4$ have demonstrated no long-lived electronegativity~\cite{Hunter1986a, Hunter1986}. 

\section{Background}
\label{Sec:Background}

In an ideal electron swarm, the individual electron energies~($\varepsilon$) follow a Maxwell-Boltzmann distribution. As energy is supplied from the electric field, the drifting electrons gain kinetic energy which can be dissipated through scattering interactions with the fill gas. For noble gases, there exist few low-lying excitations to remove this energy via inelastic collisions and thus the average instantaneous electron energy~($\langle\varepsilon\rangle$) rises rapidly with the electric field. Molecular fill gases typically exhibit low-lying excitations that may effectively absorb this energy and thus keep electron energies low until the energy supplied by the field between scattering events exceeds the energy lost in a collision. The $\langle\varepsilon\rangle$ value determines the possible reactions available to electrons, including DEA. It is a constant for a given gas mixture and reduced field~($E/N$) as the energy for an increased field is offset by the number of collisions per unit distance in higher pressures. Reduced fields throughout this work are presented in units of Townsend~(1~Townsend~(Td)=1$\times10^{-17}~Vcm^2$). 

% To determine the reduced field, $E/N$, in units of Townsend~(Td) from observable detector properties, Equation~\ref{eq:EonN} is used. 

% \begin{equation}
%     \frac{E}{N} = \frac{E\niss{drift}}{\frac{P}{P_0} n_0},
%     \label{eq:EonN}
% \end{equation}

% where $E$ is the electric field within the drift region, $P$ is the detector pressure in Torr, $n_0$ is the Loschmidt number~($\approx$2.686${\times}10^{19}$cm$^{-3}$).  conversion as 1~Td = $1{\times}10^{-17}~$Vcm$^{2}$. 

\begin{figure}[]%% placement specifier
\centering%% For centre alignment of image.
% This file was created with tikzplotlib v0.10.1.
\begin{tikzpicture}

\definecolor{crimson2143940}{RGB}{214,39,40}
\definecolor{darkgray176}{RGB}{176,176,176}
\definecolor{darkorange25512714}{RGB}{255,127,14}
\definecolor{forestgreen4416044}{RGB}{44,160,44}
\definecolor{mediumpurple148103189}{RGB}{148,103,189}
\definecolor{sienna1408675}{RGB}{140,86,75}
\definecolor{steelblue31119180}{RGB}{31,119,180}

\begin{axis}[
log basis x={10},
log basis y={10},
tick align=outside,
tick pos=left,
x grid style={darkgray176},
xlabel={Reduced Field (Td)},
xmajorgrids,
xmin=0.064231061301388, xmax=1089.81540366495,
xmode=log,
xtick style={color=black},
y grid style={darkgray176},
ylabel={Average Electron Energy \(\displaystyle \langle \epsilon\)\(\displaystyle \rangle\) (eV)},
ymajorgrids,
ymin=0.0462284978046909, ymax=25.9341760371531,
ymode=log,
ytick style={color=black}
]
\addplot [semithick, steelblue31119180]
table {%
0.1 0.06164
0.1094 0.06208
0.1196 0.06254
0.1308 0.06304
0.143 0.06358
0.1564 0.06415
0.171 0.06476
0.187 0.06542
0.2045 0.06612
0.2236 0.06686
0.2446 0.06765
0.2674 0.06849
0.2925 0.06938
0.3198 0.07031
0.3497 0.0713
0.3825 0.07234
0.4182 0.07342
0.4574 0.07456
0.5002 0.07575
0.547 0.07698
0.5981 0.07826
0.6541 0.07958
0.7153 0.08094
0.7822 0.08234
0.8554 0.08378
0.9354 0.08525
1.023 0.08675
1.119 0.08829
1.223 0.08985
1.338 0.09145
1.463 0.09307
1.6 0.09473
1.749 0.09646
1.913 0.09821
2.092 0.1
2.288 0.1019
2.502 0.1038
2.736 0.1058
2.992 0.1079
3.271 0.1101
3.577 0.1125
3.912 0.115
4.278 0.1178
4.678 0.1207
5.116 0.1239
5.595 0.1275
6.118 0.1315
6.69 0.1361
7.316 0.1413
8.001 0.1474
8.749 0.1547
9.568 0.1631
10.46 0.174
11.44 0.1866
12.51 0.2028
13.68 0.2224
14.96 0.2459
16.36 0.2773
17.89 0.3199
19.57 0.3776
21.4 0.461
23.4 0.5844
25.59 0.7613
27.98 0.9983
30.6 1.289
33.46 1.617
36.59 1.963
40.02 2.313
43.76 2.662
47.85 3.012
52.33 3.369
57.23 3.739
62.58 4.124
68.43 4.525
74.84 4.936
81.84 5.351
89.49 5.761
97.87 6.162
107 6.551
117 6.929
128 7.296
140 7.657
153 8.015
167.4 8.374
183 8.741
200.1 9.12
218.9 9.517
239.3 9.937
261.7 10.39
286.2 10.88
313 11.41
342.3 11.99
374.3 12.63
409.3 13.34
447.6 14.12
489.5 14.99
535.3 15.95
585.4 17.01
640.1 18.17
700 19.45
};
\addplot [semithick, darkorange25512714, dashed]
table {%
0.064231061301388 0.16
1089.81540366495 0.16
};
\addplot [semithick, sienna1408675, dashed]
table {%
0.064231061301388 20
1089.81540366495 20
};
\addplot [semithick, mediumpurple148103189, dashed]
table {%
0.064231061301388 12.5
1089.81540366495 12.5
};
\addplot [semithick, forestgreen4416044, dashed]
table {%
0.064231061301388 5
1089.81540366495 5
};
\addplot [semithick, crimson2143940, dashed]
table {%
0.064231061301388 5.9
1089.81540366495 5.9
};
\draw (axis cs:0.1,0.15) node[
  scale=0.8,
  fill=white,
  draw=black,
  line width=0.4pt,
  inner sep=3pt,
  anchor=base west,
  text=black,
  rotate=0.0
]{Excitation};
\draw (axis cs:0.1,5) node[
  scale=0.8,
  fill=white,
  draw=black,
  line width=0.4pt,
  inner sep=3pt,
  anchor=base west,
  text=black,
  rotate=0.0
]{DEA};
\draw (axis cs:0.1,11.2) node[
  scale=0.8,
  fill=white,
  draw=black,
  line width=0.4pt,
  inner sep=3pt,
  anchor=base west,
  text=black,
  rotate=0.0
]{Excitation};
\draw (axis cs:0.1,18.2) node[
  scale=0.8,
  fill=white,
  draw=black,
  line width=0.4pt,
  inner sep=3pt,
  anchor=base west,
  text=black,
  rotate=0.0
]{Ionization};
\end{axis}

\end{tikzpicture}
\caption{Simulation using BOLSIG+~\cite{pancheshnyi2012lxcat, Hagelaar2005} of electron energies in CF$_4$ using cross section data from Ref~\cite{TRINITI2026}. The rise in energies between 15-80~Td corresponds to the transition from low lying vibrational excitations to DEA, where very little energy can be transferred by inelastic collisions.}\label{fig:BOLSIGCF4}
\end{figure}
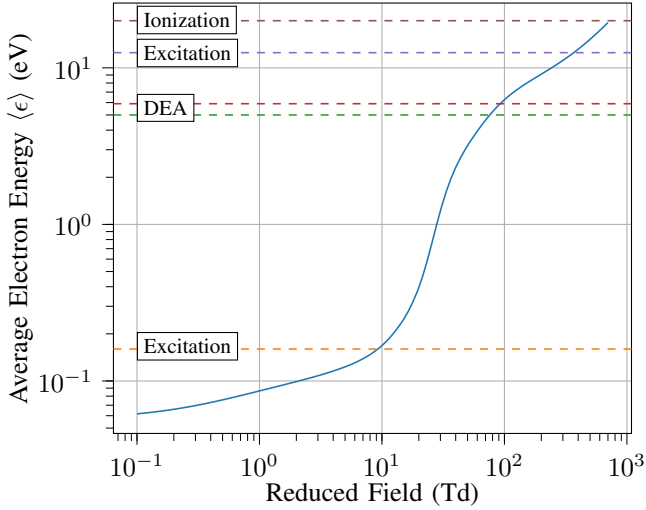

Molecular fill gases such as CF$_4$ exhibit low-energy vibrational excitations, and are often used to lower $\langle\varepsilon\rangle$ values within a gas mixture through inelastic collisions. An example of  $\langle\varepsilon\rangle$ versus reduced field for CF$_4$ is shown in Figure~\ref{fig:BOLSIGCF4}. This effect is observable within detector conditions as both an increase in drift velocity and a lowering of electron diffusion~\cite{Pansky1995}. Despite this, the observable signal diffusion within CF$_4$ is suprathermal under typical operating conditions and could be improved with NID. As CF$_4$ possesses an electron affinity of $-0.7$~eV~\cite{Christophorou1996}, it does not form a stable negative ion and is not considered an electronegative gas. Above the DEA threshold, the electron energy is sufficient to sever the CF$_3$-F bond, which has a bond dissociation energy of 5.6~eV~\cite{Macneil1970}. Both F$^-$ and CF$_3^-$ negative ions may be produced via~\cite{Spyrou1982}: 

\begin{align}
    \mathrm{CF}_4^{-*} &\rightarrow \mathrm{CF}_3^{-} + \mathrm{F} \label{eg:cf3}\mathrm{, and}\\
     &\rightarrow \mathrm{F}^{-} + \mathrm{CF}_3. \label{eg:f}
% & = F_1 * F_2
\end{align}

Due to the increased electron affinity of F$^-$ compared to CF$_3^-$~(3.45 and 1.8~eV respectively~\cite{Macneil1970}), the production of F$^-$ occurs at lower energies~\cite{Christophorou1996,Macneil1970}, with a higher interaction cross section. However, there is an overlapping energy region where it is possible to produce two different ion species from DEA interactions with CF$_4$. 

% \begin{figure}[]%% placement specifier
% \centering%% For centre alignment of image.
% % \includegraphics[width=0.5\textwidth]{Images/Old_images/Reaction_rates.pdf}
% \input{Figures/Reaction_rates}
% \caption{BOLSIG+ simulation of reaction rates for various CF$_4$ scattering interactions. Data from Ref~\cite{TRINITI2026}. \label{fig:BOLSIGreactions}}
% \end{figure}

Depending on the chosen swarm model, the CF$_4$ DEA threshold is expected to occur around 12-25~Td~\cite{TRINITI2026,Kurihara2000}. Previous experimental studies conducted at pressures far below particle detector operating conditions indicate a range for DEA in CF$_4$ from 30-80~Td~\cite{Hunter1986, JDutton1987a, Anderson1992}. These reduced fields are large compared to the conventional drift fields applied to gas TPCs, which tend to operate below 20~Td in order to minimise effective diffusion~\cite{Caldwell2009, Deaconu2017}. The value of $\langle \varepsilon \rangle$ required for DEA~(2.15~eV~\cite{Macneil1970}) is also significantly less than that required for ionization within the detector~(14.7~eV~\cite{Christophorou1996}), thus excluding the possibility of avalanche until reduced fields exceed 100~Td.

% \begin{figure}[]%% placement specifier
% \centering%% For centre alignment of image.
% \includegraphics[width=0.5\textwidth]{Images/Townsend.pdf}
% \caption{Townsend coefficients for interactions in pure CF$_4$. Data from Ref~\cite{TRINITI}}\label{fig:BOLSIGalpha}
% \end{figure}

% #########################################################################################################

\section{Garfield++ Simulation}
\label{Sec:Simulation}

 \begin{center}
\begin{figure*}[!h]
\centering

\subfloat[]{\includegraphics[height=7cm]{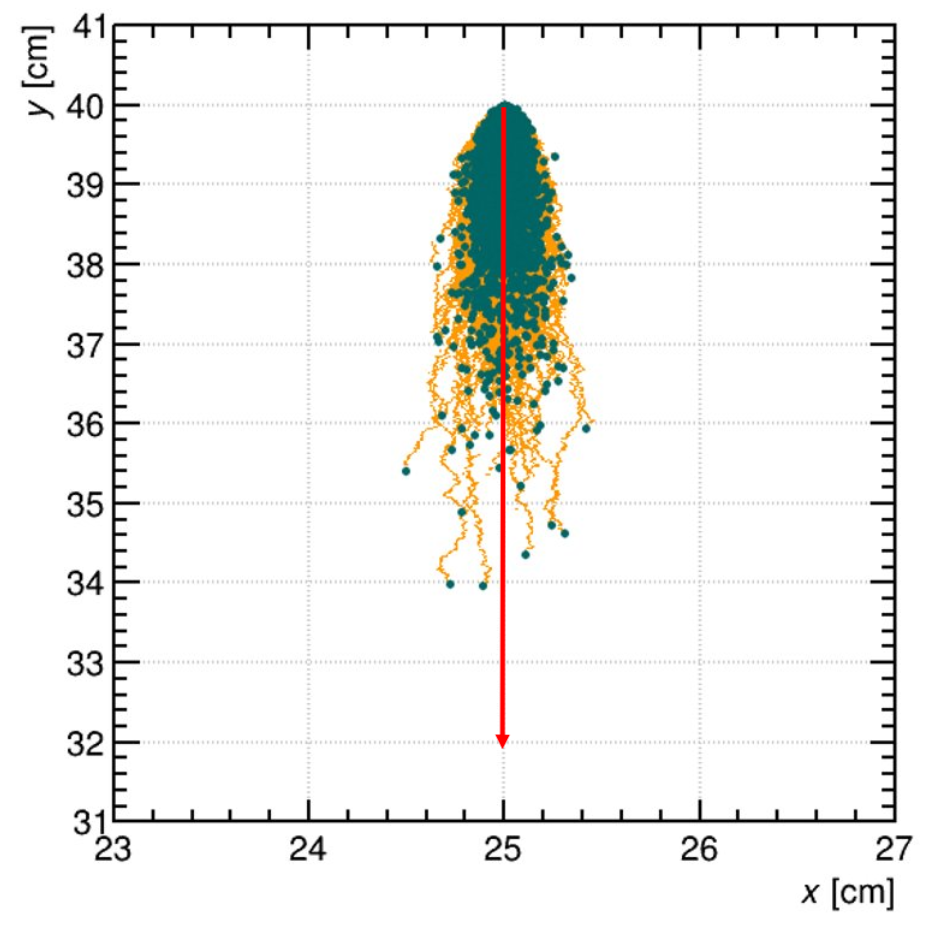}} ~
\subfloat[]{\includegraphics[height=7cm]{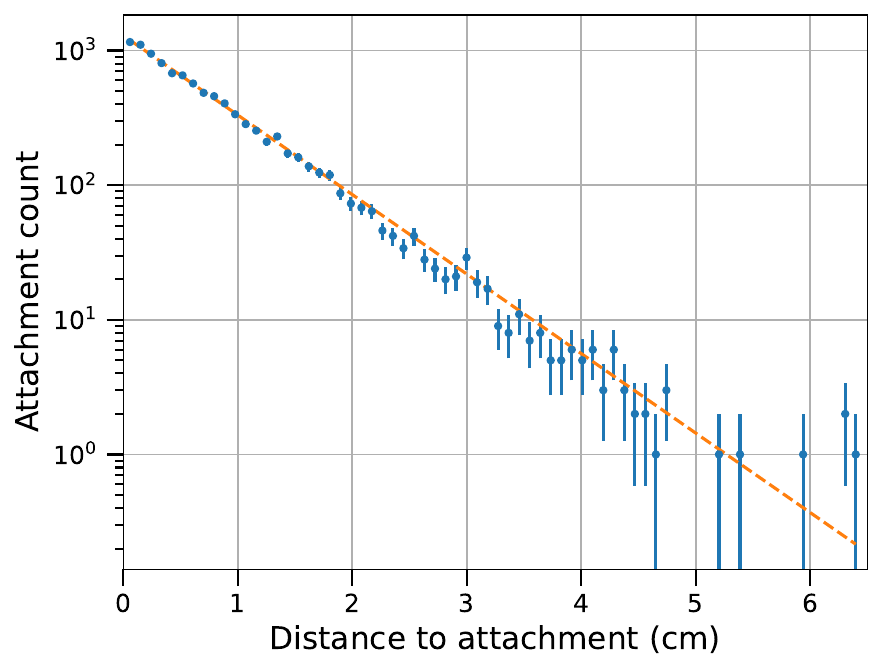}}

\caption[]{Simulated dissociative electron attachment in 38~Torr CF$_4$ at 48~Td. (a)~10000 electron drift tracks in Garfield++, beginning at (25,40) with the drift direction indicated by the red arrow and attachment locations in green. (b)~The attachment locations with respect to the drift direction~($\Delta y$), fitted with an exponential (orange dashed line).}
\label{fig:GarfieldAttach}
\end{figure*}
\end{center}

A simplified detector geometry was simulated using the Garfield++ simulation framework~\cite{Veenhof2020} for the purposes of modelling the expected DEA behaviour in CF$_4$. The simulation consisted of a 50~cm long gas volume located within two 50~cm parallel planes. Electrons were placed 40~cm from the readout plane and drifted in a pure CF$_4$ gas environment. Pressures were varied between 38 and 760~Torr, and voltages were simulated to create reduced fields between 0.1 and 50~Td. Efforts were made to extend the simulation above 50~Td, however occasional avalanche ionisation processes introduced computational difficulties. Thus, the reduced field was limited to 50~Td. 

The CF$_4$ scattering cross sections used in the Garfield++ model indicate a DEA threshold of 20~Td. Below this value, the drifted electrons reach the anode plane of the detector. Above this DEA threshold, however, electrons underwent attachment and the simulation recorded attachment locations. These locations could then be fitted to determine attachment properties of the gas.

\subsection{Attachment Length}

\begin{figure}[]%% placement specifier
%% Use \includegraphics command to insert graphic files. Place graphics files in 
%% working directory.
\centering%% For centre alignment of image.
\includegraphics[width=\columnwidth]{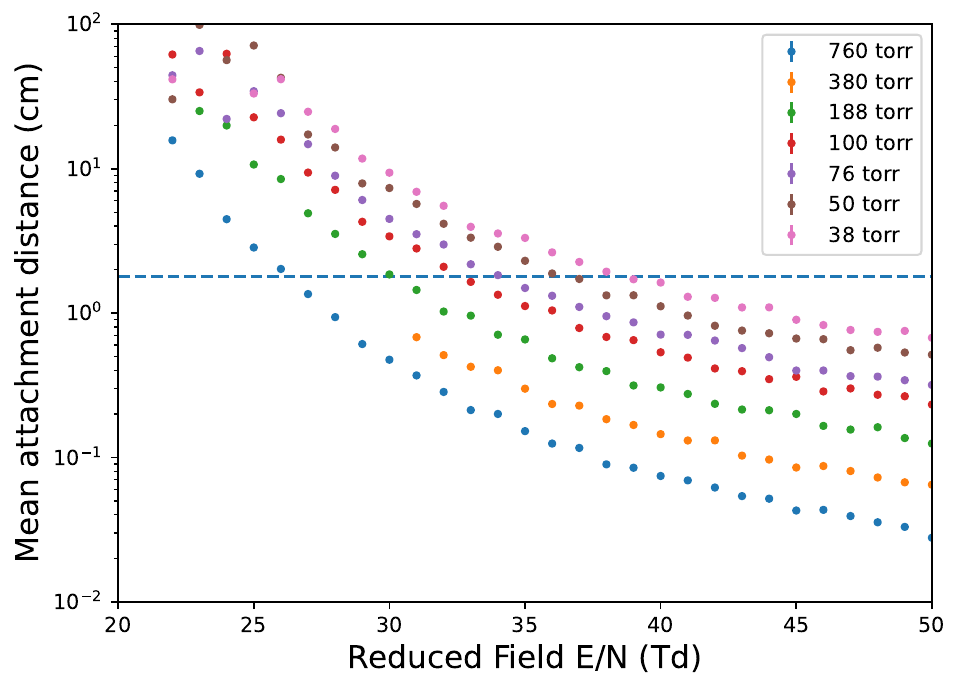}
%% Use \caption command for figure caption and label.
\caption{Simulated mean electron attachment distances in pure CF$_4$. The attachment length grows shorter with increasing reduced field. The dashed horizontal line indicates the position of the source above the readout plane used in the experimental aspect of this work.}\label{fig:Garfield}
\end{figure}

 Figure~\ref{fig:GarfieldAttach} shows an example of these simulations. The distribution of attachment locations with respect to the drift direction was fitted with an exponential function, where the exponent of the fit is the attachment coefficient ($\eta$). The reciprocal of the attachment coefficient is the mean attachment length ($\bar{\eta}$), which corresponds to the average electron displacement in the drift direction before forming a negative ion~(see Figure~\ref{fig:Garfield}). For most detection applications, $\bar{\eta}$ would ideally be as small as possible to maintain the position resolution of the original track.

\begin{figure}[]%% placement specifier
%% Use \includegraphics command to insert graphic files. Place graphics files in 
%% working directory.
\centering%% For centre alignment of image.
\includegraphics[width=\columnwidth]{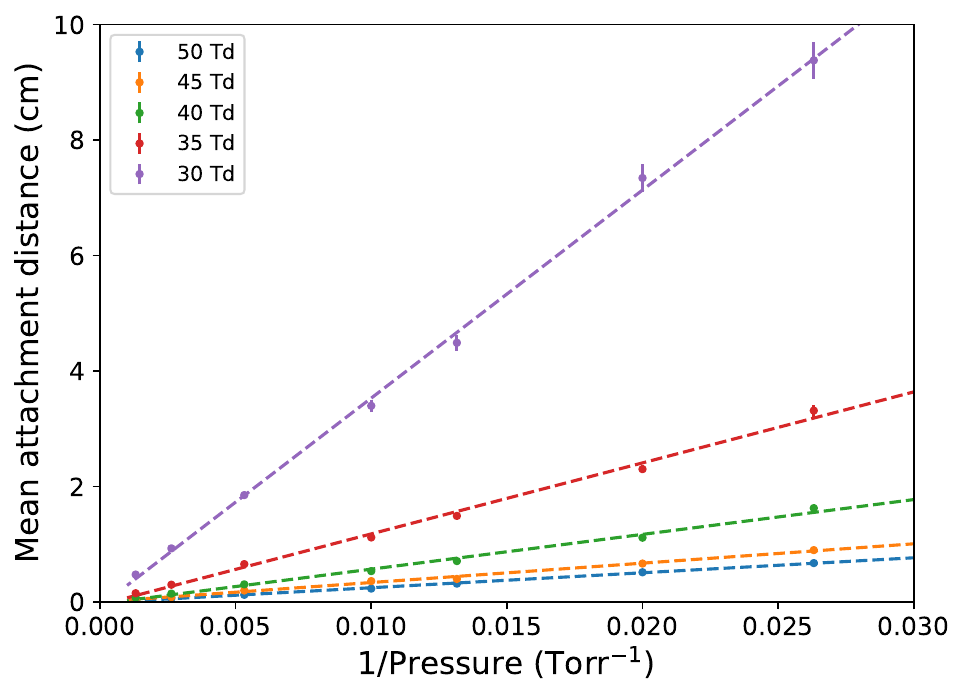}
%% Use \caption command for figure caption and label.
\caption{Simulated mean attachment lengths as a function of inverse pressure for simulated dissociative electron attachment in CF$_4$. The linear dependence is consistent with a two-body interaction.}\label{fig:Garfield1onp}
\end{figure}

 The simulated attachment lengths within the region modelled tended to shorten with larger reduced fields, and may further shorten at higher fields, however these higher fields could not be simulated due to ionization above 50~Td as discussed above. The results also indicate a pressure dependence, with higher gas pressures leading to smaller attachment lengths for a given reduced field. This is expected; higher pressures imply more gas collisions per unit drift distance. The linear pressure dependence shown in Figure~\ref{fig:Garfield1onp} is indicative of a purely two-body interaction as expected from Equations~\ref{eg:cf3} and~\ref{eg:f}.

 The simulations predict attachment lengths as short as 200~$\mu$m in 760~Torr pure CF$_4$. Such distances are comparable to the spatial pitch of readout technologies being considered for directional rare-event physics detectors~\cite{Vahsen2020, Miuchi2011, Mazzitelli2023}. Such an experimental setup would present challenges, requiring drift fields (E$\niss{drift}$) as great as 13450~V/cm. However, modest attachment lengths on the order of 1~cm are predicted to be possible in pressures below 50~Torr with E$\niss{drift}$ no greater than 1~kV/cm. An initial experimental investigation was therefore undertaken at low pressures to demonstrate DEA in pure CF$_4$.

% ########################################################################################################

\section{Experimental Setup}
\label{Sec:Setup}
\begin{figure}[]%% placement specifier
%% Use \includegraphics command to insert graphic files. Place graphics files in 
%% working directory.
\centering%% For centre alignment of image.
\includegraphics[width=\columnwidth]{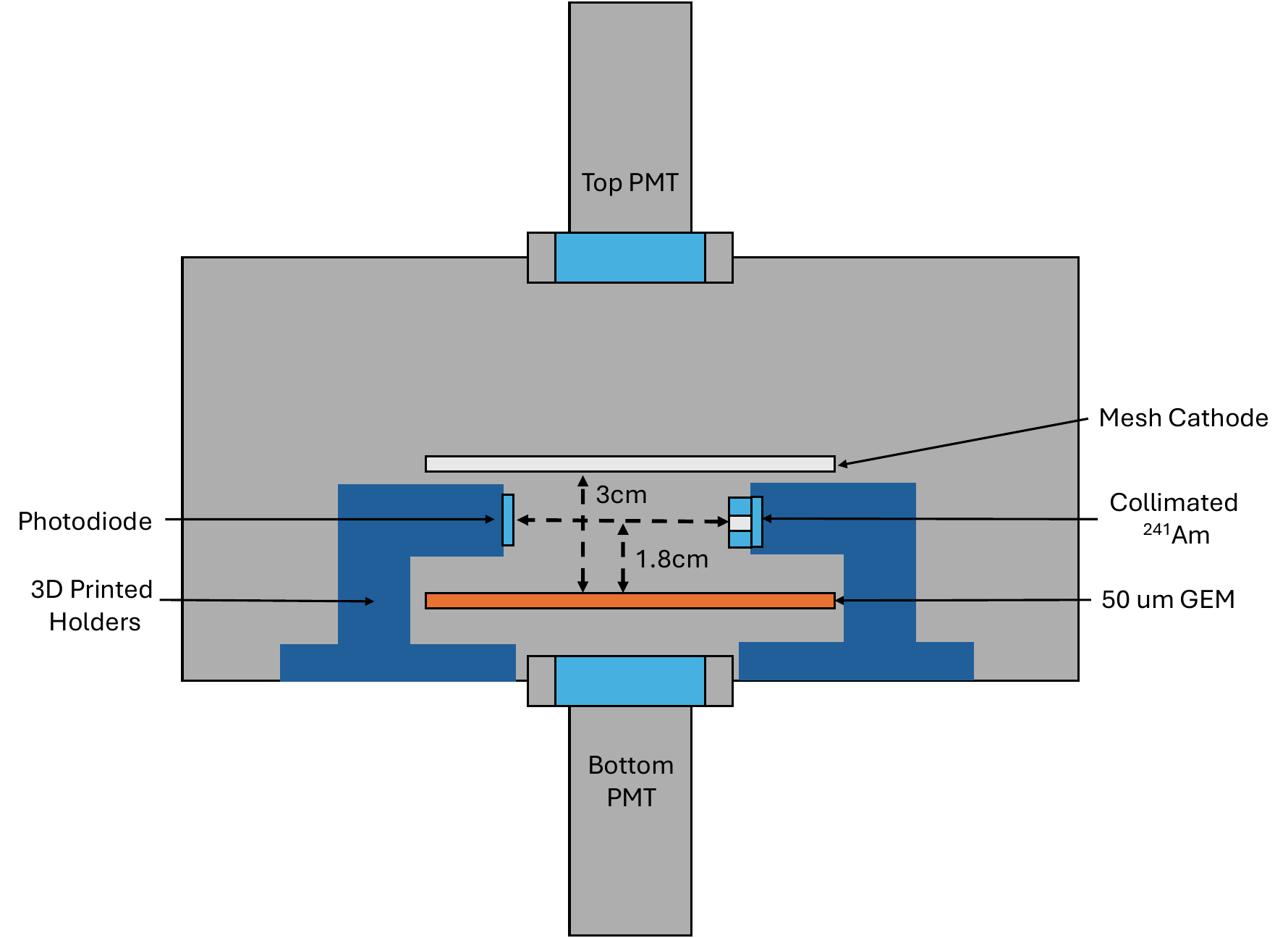}
%% Use \caption command for figure caption and label.
\caption{The CYGNUS-n experimental setup used for the measurements described in this work.}\label{fig:CYGndiag}
\end{figure}

Measurements were performed in a small TPC consisting of a 3~cm drift region between a stainless steel mesh cathode and a Gas Electron Multiplier~(GEM) amplification stage. As CF$_4$ is a scintillating gas, optical measurements were recorded using two 1.5~inch Hamamatsu R9420-100 Photomultiplier Tubes~(PMTs) observing the light through windowed flanges. One PMT was located above the cathode mesh and the other was positioned directly below the base of the GEM. A diagram of the experimental setup is shown in Figure~\ref{fig:CYGndiag}.

A standard CERN 100~mm$\times$100~mm GEM -- 50~$\mu$m thickness, with a hole diameter of 70~$\mu$m and a pitch of 140~$\mu$m -- was used as the avalanche gain stage. The operational GEM voltage~($\Delta$V$\niss{GEM}$) was maintained at a fixed potential of 280~V for all measurements by applying a positive high voltage to the bottom GEM layer using an iseg EHS 80 60p power supply. The top of the GEM was maintained at a ground potential through a 5~M$\Omega$ quenching resistor in series. The GEM voltage was chosen by incrementally increasing $\Delta$V$\niss{GEM}$ in 30~Torr CF$_4$, with a drift field ($E\niss{drift}$) of 200~V/cm, until electron avalanche scintillation light gave an observable PMT signal greater than 100~mV on both PMTs. The fixed $\Delta$V$\niss{GEM}$ value resulted in different gas gains between measurements at different pressures. No effort was made to maximise gain due to concerns about destructive sparking in low pressure~($<$~100~Torr) gases with a 50~$\mu$m thin GEM~\cite{Phana}.

Adjustments of E$\niss{drift}$ were made by altering the negative bias applied to the mesh cathode. The voltage was supplied using an iseg EHS 80 60n through a 60~M$\Omega$ resistance to prevent a current trip of the power supply in the event of occasional sparking. The high voltage supplies were monitored to ensure no measurable current~($<$ $\mu$A) was drawn during normal operation of the detector. 

To minimise the effects of contamination from gas impurities and material outgassing, the startup procedure consisted of evacuating the detector and gas system to $\mu$Torr pressures overnight using an Edwards nXDS10i scroll pump and maintaining this vacuum until measurements were ready to commence. The detector was then back-filled to operational pressures with analytical grade CF$_4$~(99.999\% purity, COREGAS) via an Alicat MCS-200SCCM Mass Flow Controller~(MFC). When at the desired operating pressure, experimental measurements proceeded with the detector in a constant flow configuration at an MFC input rate of 10~SCCM. The pressure was maintained using an Alicat pressure controller backed by a Pfeiffer MVP 015-4 diaphragm pump.

Particle events were generated within the detector using a collimated 3.6~MBq $^{241}$Am source. The source projected 5.4~MeV $\alpha$ particles parallel to, and 1.8~cm above, the GEM plane, towards a 10$\times$10~mm silicon photodiode~(API PDB-C613-2) located 6~cm from the source. In the low pressures studied, incomplete energy deposition was expected in the gas so that the $\alpha$ particles reached the photodiode, which acted as the trigger. Geant4 simulations indicate that alpha particles crossing the TPC in 50~Torr gas deposited approximately 1~MeV in the region above the GEM. Measurements were made in a pressure range of 15-50~Torr. Measurements above 50~Torr became impractical due to the straggling and stopping of the alpha particles in the gas, reducing the photodiode trigger rate.

Signals from the top and bottom PMTs, the base of the GEM (via a Cremat CR-110 preamplifier), and the photodiode were recorded using a 16~bit, 125~MSPS  digitiser (CAEN 2740). In order to reduce data volumes, the acquisition length was chosen to be the shortest period judged to be sufficient to collect signals of interest. Typical recorded waveforms are 100-400~$\mu$s in length. Given the unknown nature of negative ion species and the possibility of multiple ion species, longer waveform acquisitions of as long as 2~ms were used to search for additional delayed signals, with no indication of ion arrivals at these times.

 For this investigation, a range of reduced fields at pressures between 15 and 50~Torr were studied, starting at reduced fields near 1~Td and increasing the drift field incrementally through the entire range of the 6~kV power supply. Measurements at the maximum 6~kV voltage exhibited evidence of sparking, likely from the cathode to photodiode, so were not included in the analysis.

%#############################################################################################

%\begin{center}
\begin{figure*}[h]
\centering
\includegraphics[width=0.75\linewidth]{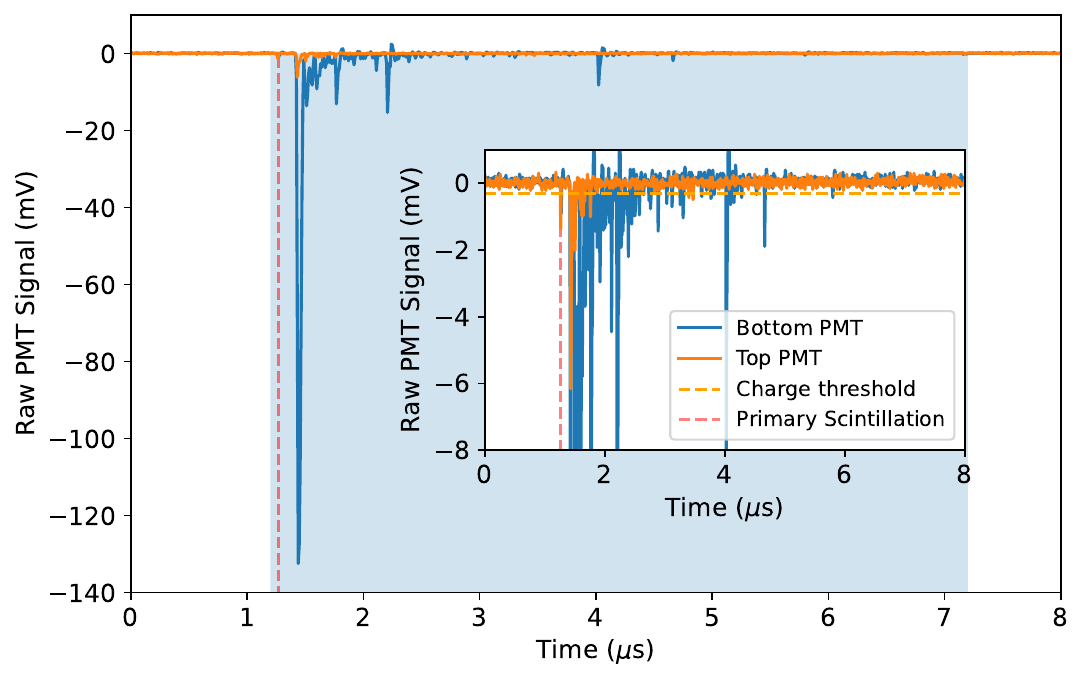} 
\caption[]{Baseline subtracted PMT waveforms from the top~(orange) and bottom~(blue) PMTs in the electron avalanche region for an event with a drift field of 200~V/cm, in 35~Torr CF$_4$~(17.5~Td). The top PMT saw more primary scintillation (vertical dashed red line) on average, while the bottom PMT saw a stronger avalanche light signal. Integrated signal intensity from the electron avalanche region~(shaded blue region) extends for 6~$\mu$s. Segments of the signal exceeding the charge threshold~(horizontal dashed orange in inset) within this region are integrated to determine the signal intensity within the electron avalanche. \label{fig:Avalanche}}
\end{figure*}
%\end{center}

\section{Signal analysis and results}
\label{Sec:Analysis}

Raw PMT signals taken at low drift fields well below the expected DEA threshold are shown in Figure~\ref{fig:Avalanche}. The measurements are characteristic of conventional electron drift, with a prompt primary scintillation signal followed by an electron avalanche from primary electrons reaching the GEM plane. We attribute the late pulses evident in the first few $\mu$s after the arrival of the avalanche light to PMT afterpulsing. The sub~200~ns time difference between the prompt peak and the avalanche signal is consistent with the drift velocity for electrons in CF$_4$ traversing the 1.8~cm drift volume for these reduced fields~\cite{Christophorou1996}. 

The analysis outlined in this work focuses on the optical signals from the PMTs, primarily the bottom PMT. The top and bottom PMTs measure similar amounts of primary scintillation, while significantly more avalanche scintillation was observed on the bottom PMT, consistent with the measurement geometry. All analysis of the avalanche signals therefore exclusively focused on the bottom PMT. 

\begin{figure*}[!h]
\centering
\subfloat[Below DEA threshold]{\includegraphics[width=\textwidth]{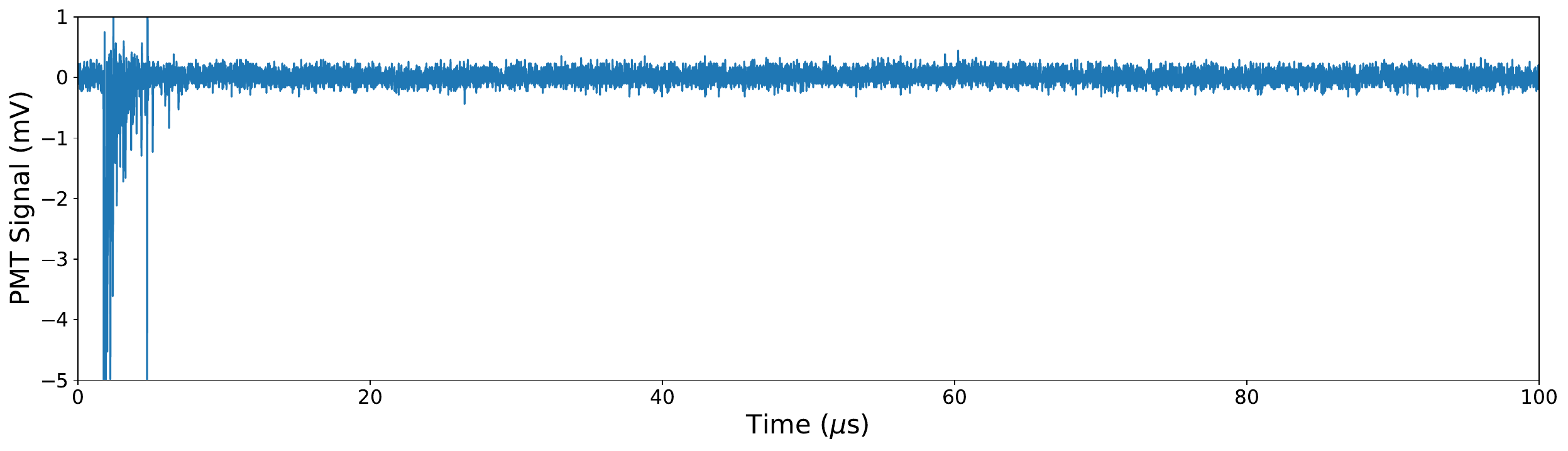}} \\
\subfloat[Above DEA threshold]{\includegraphics[width=\textwidth]{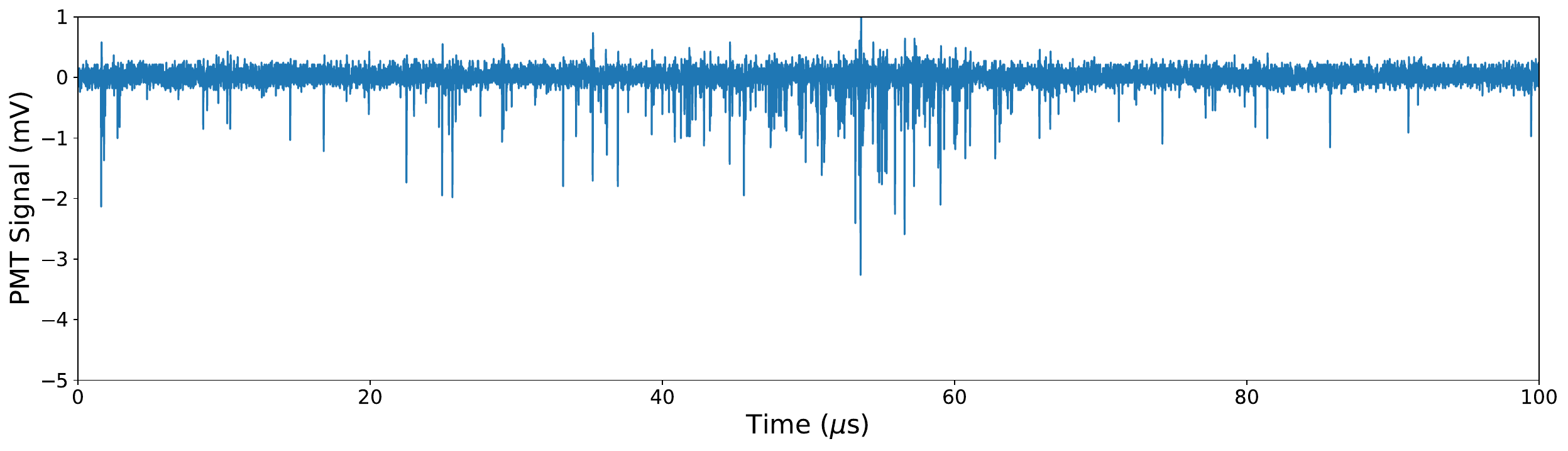}}
\caption[]{Baseline subtracted signals from the bottom PMT. Both signals are recorded in 20~Torr pure CF$_4$ with (a) 100~V/cm drift field (14 Td) and (b) 1600~V/cm drift field (226 Td). \label{fig:Abovenbelow}}
\end{figure*}

Examples of the acquired traces for bottom PMT signals taken with drift fields below and above the expected DEA threshold are presented in Figure~\ref{fig:Abovenbelow}. At the low drift field, the fast electron signal is readily observed. In contrast, the higher drift field measurement exhibits a highly suppressed electron avalanche signal. Moreover, late scintillation signals, out to 250~$\mu$s from the trigger region are seen. These late signals are absent from the low field measurements. The late signals are distributed in a broad time peak, consistent with a low-inclination alpha particle track. The location of the peak was seen to move as the reduced field was varied, as discussed below. We attribute these late scintillation signals to the transfer of primary electrons to low-mobility negative ions via DEA, and we will refer to the signals in this region as negative ions hereafter. 

%\begin{center}
%\end{center}

\subsection{Signal processing}

The baseline of each PMT signal was determined using the mean of the first 150~samples~(1.2~$\mu$s) of the waveform, and then subtracted from the waveform. This region preceded the arrival of the primary scintillation light at 1.45~$\mu$s after the start of the waveform. The baseline distribution was typically offset from 0~mV, following an approximately normal distribution with a standard deviation no wider than 0.1~mV. The limited number of events with baselines outside this main distribution, $<$~1~\%, were excluded from further analysis. This baseline cut, was the first of two quality cuts made to remove spurious and unwanted events from the data prior to analysis.

% \subsubsection{Single Event Cut}

 The use of the photodiode trigger heavily suppressed unwanted backgrounds. A simple cut threshold on the photodiode signal excluded the rare instances of multiple photodiode hits within a given trigger. However, $\alpha$ particles that reached the TPC volume, which did not interact with the photodiode were also present in the data as pileup events. A cut was therefore implemented to remove alpha particle pileup by rejecting events that contained multiple electron avalanches in a single waveform. The cut was made on the magnitude of the signal intensity from single ion avalanches, which are discussed in section~\ref{sec:singleions}. Events containing clusters of photons arriving outside the electron avalanche window with a magnitude greater than 100 arbitrary charge units were excluded. Across all datasets, fewer than 10~\% of recorded events contained triggers that did not pass this cut.

\subsection{Integrated signal intensity}

The $^{241}$Am source used in this experiment exhibits some self-attenuation so that the $\alpha$ particles emitted into the TPC can not be taken as monoenergetic. Furthermore, the energy deposited by the $\alpha$ particles in the gas volume was incomplete given the requirement that it interact with the photodiode. This prohibited an absolute energy calibration of the detector and therefore gain measurements were not practical. Instead, relative comparisons of the scintillation light components were made by comparing the integrated PMT signal within fixed windows containing the electron avalanche and negative ion components of the waveform.

\begin{figure}[]%% placement specifier
\centering%% For centre alignment of image.
\includegraphics[width=\columnwidth]{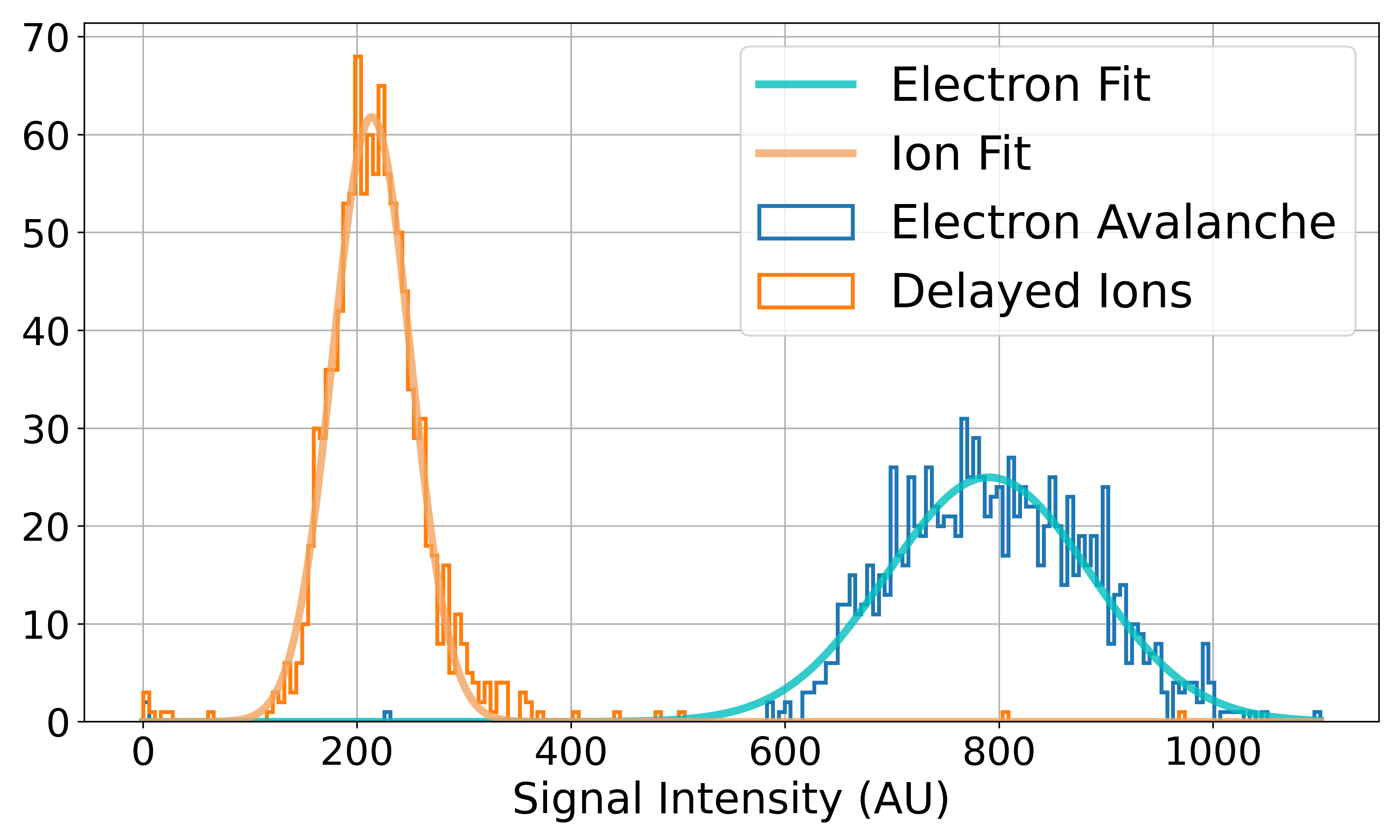}
\caption{Integrated signal intensity distributions for the primary, electron
avalanche, and negative ion regions of the PMT waveform (see text for details) at 54 Td reduced field and 45 Torr.}\label{fig:indSI}
\end{figure}

\begin{center}
\begin{figure*}[!htbp]
\centering

\subfloat{\includegraphics[width=0.5\linewidth]{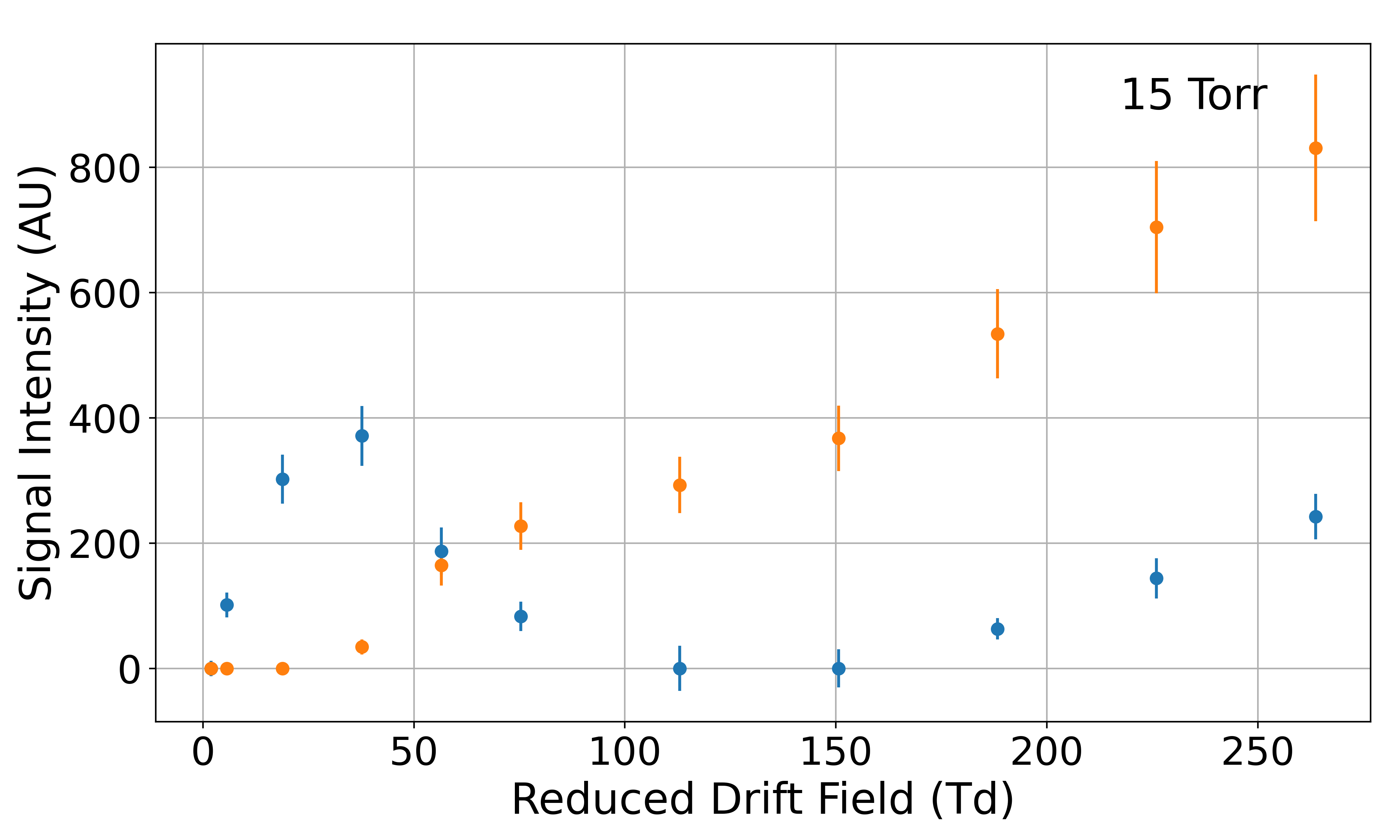}{}}
\subfloat{\includegraphics[width=0.5\linewidth]{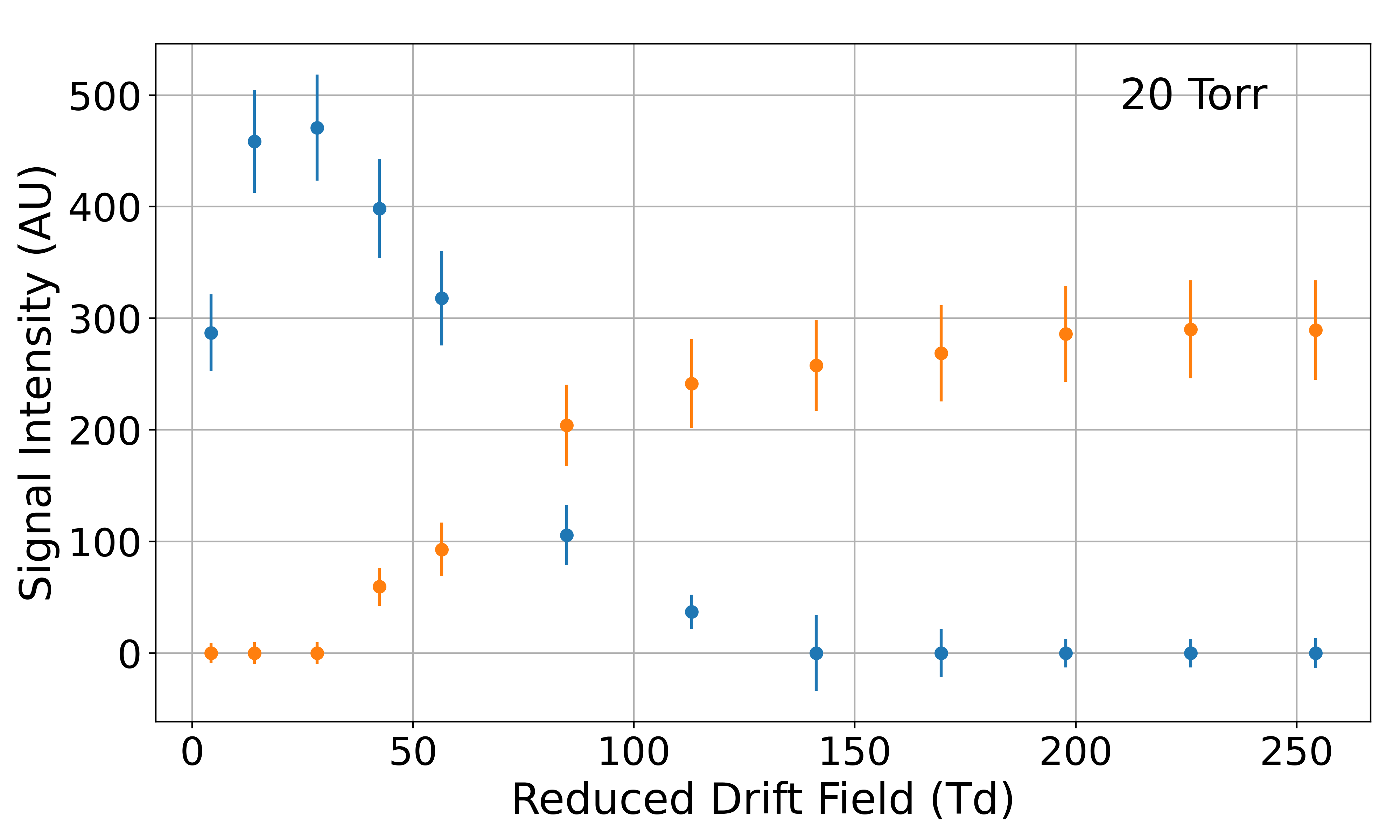}{}}\\

\subfloat{\includegraphics[width=0.5\linewidth]{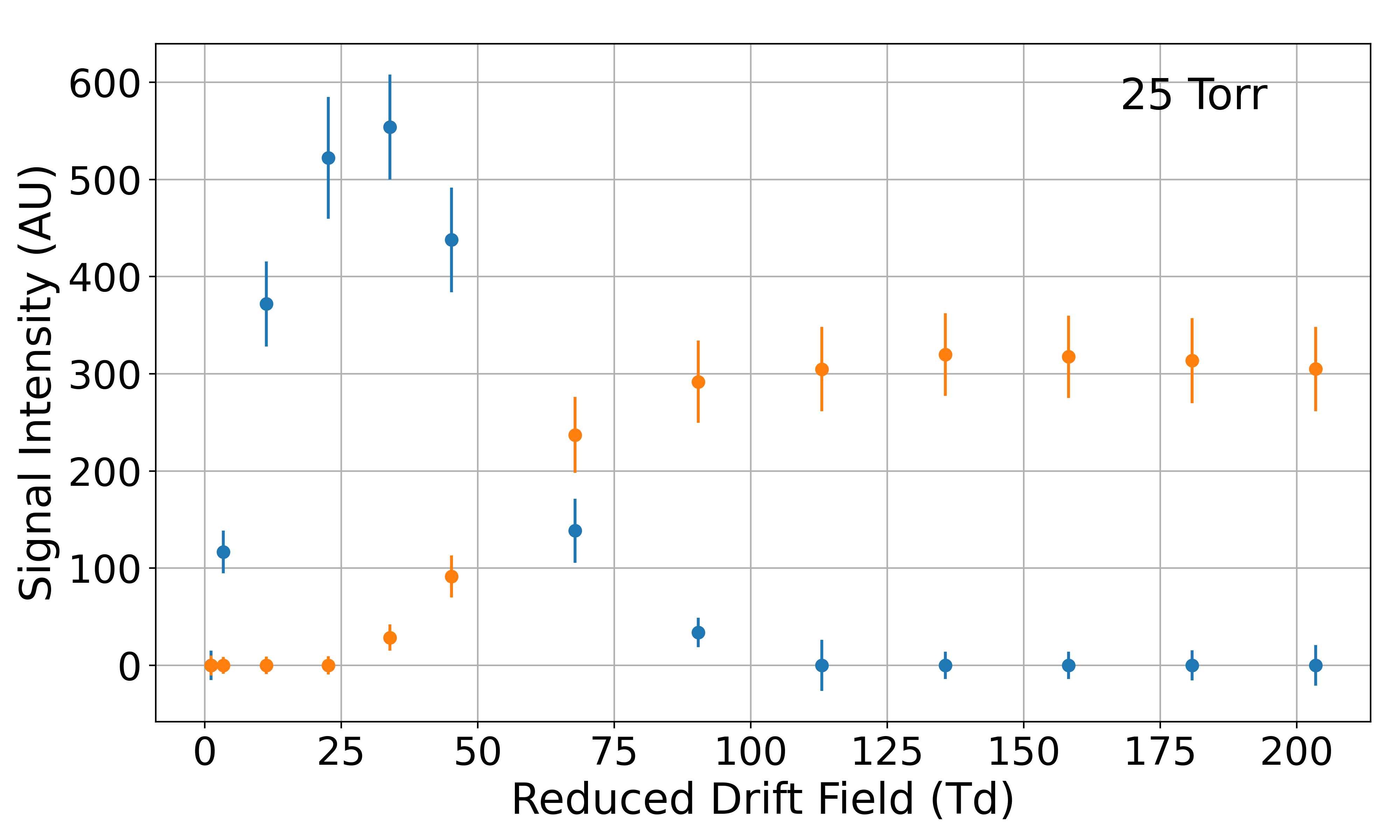}{}}
\subfloat{\includegraphics[width=0.5\linewidth]{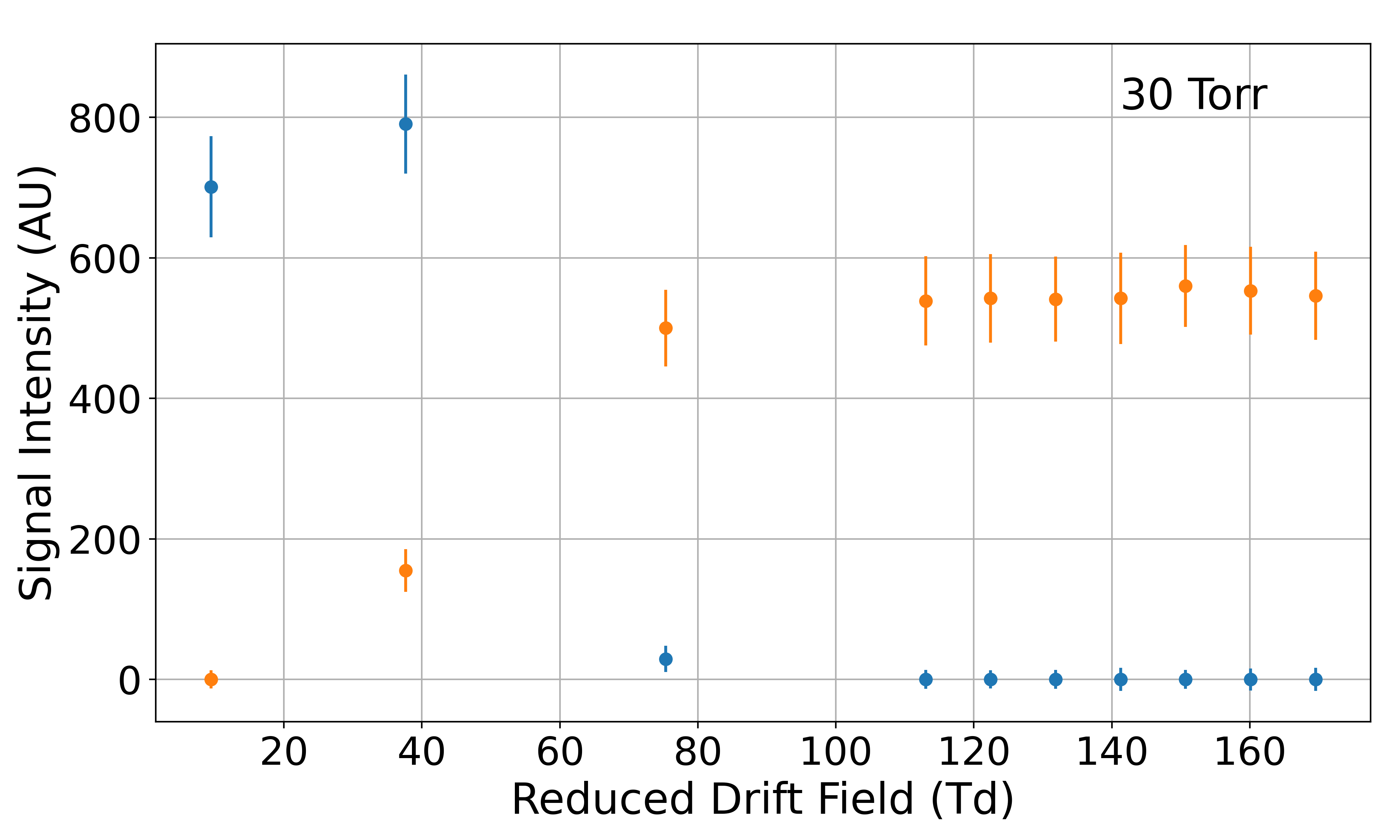}{}}\\

\subfloat{\includegraphics[width=0.5\linewidth]{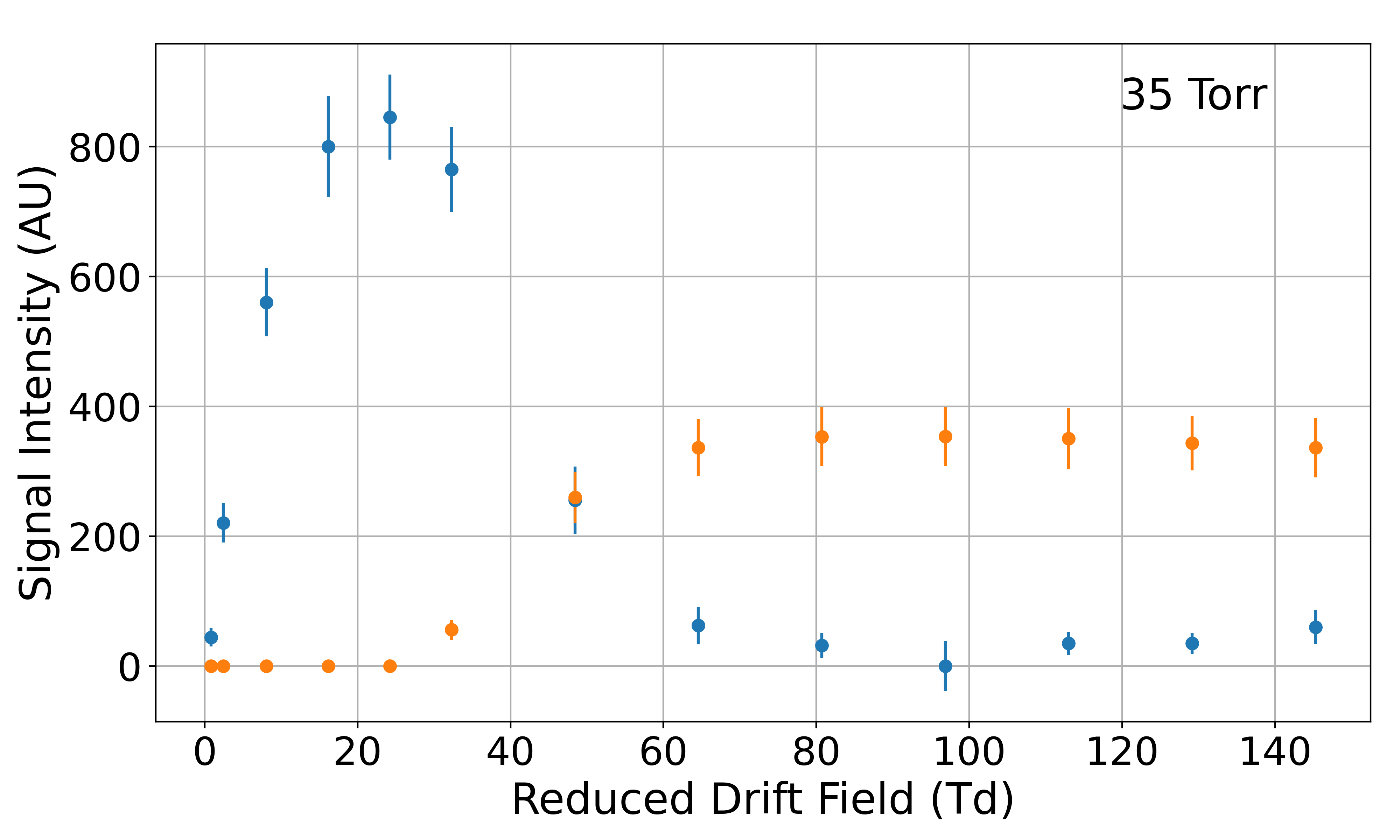}{}}
\subfloat{\includegraphics[width=0.5\linewidth]{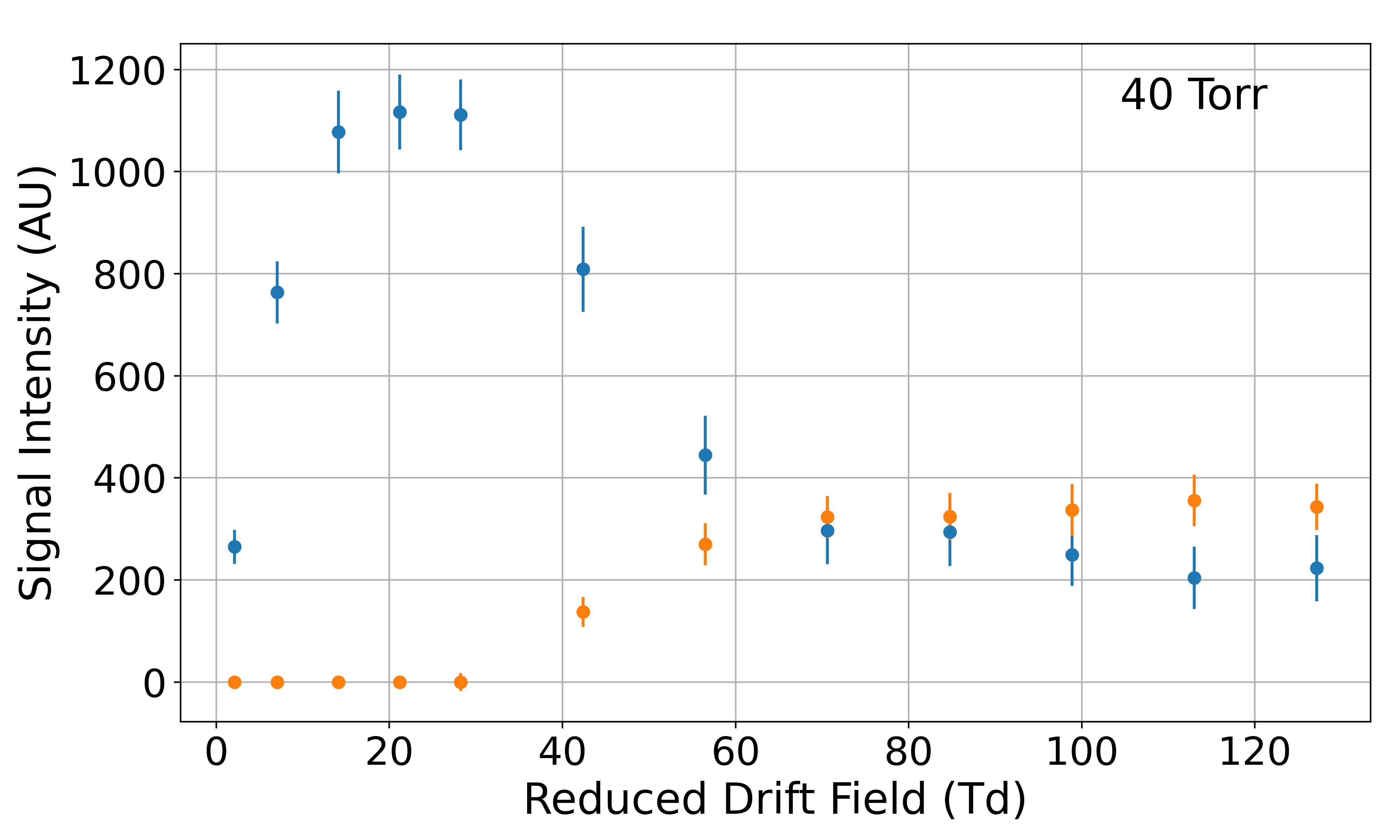}{}}\\
% \hspace*{0.25\textwidth}%
\subfloat{\includegraphics[width=0.5\linewidth]{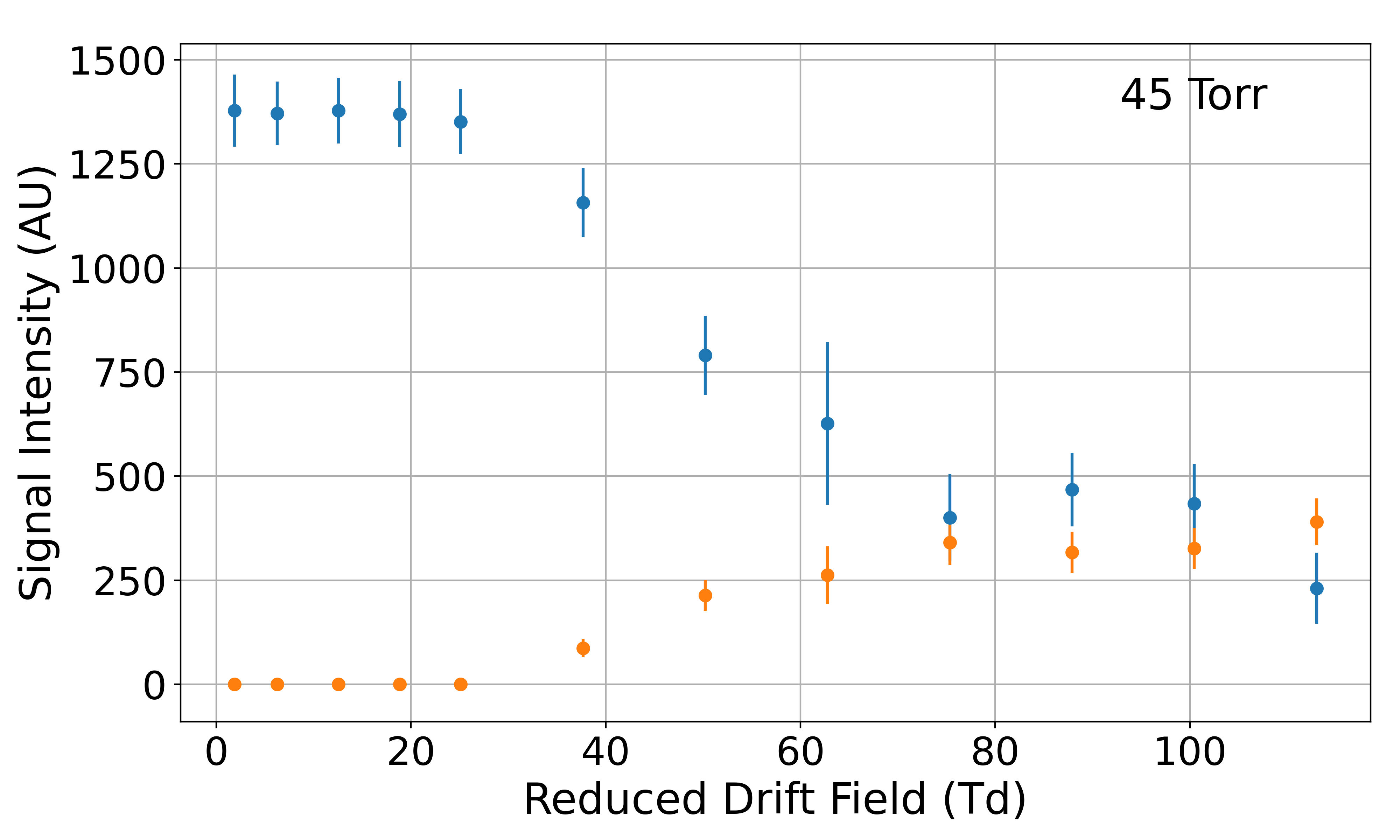}{}}
\subfloat{\includegraphics[width=0.5\linewidth]{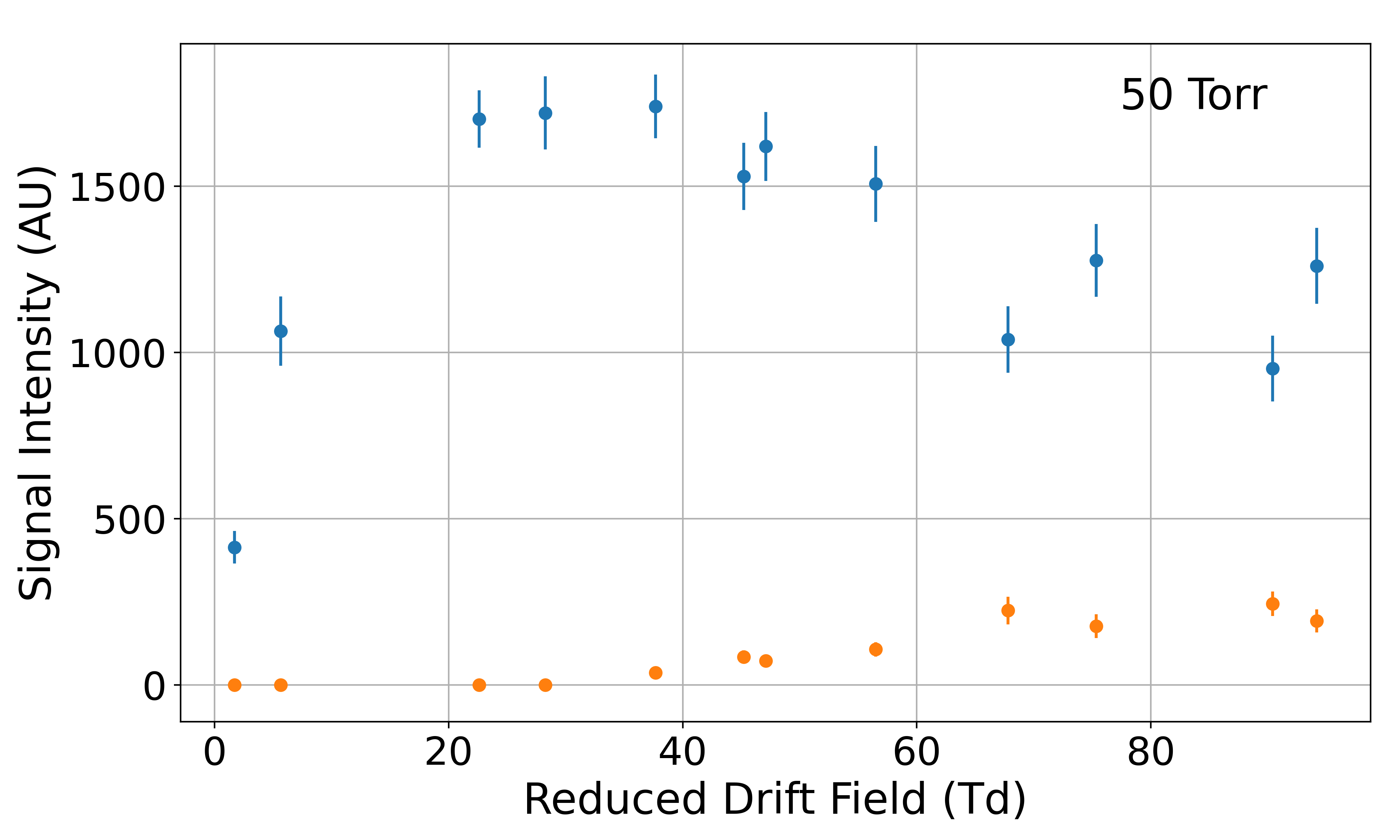}{}}\\

\caption[Evidence of DEA in signal intensity trends]{Mean values of the integrated signal intensities of electron avalanches~(blue) and negative ions region~(orange) as a function of reduced field, between 15 and 50~Torr, as indicated in the captions. Error bars represent the standard deviation of the fitted distributions. \label{fig:DEA}}
\end{figure*}
\end{center}

The 6~$\mu$s electron avalanche region spans between sample 150~(1.2~$\mu$s) from the beginning of the waveform to sample 900~(7.2~$\mu$s). As indicated in Figure~\ref{fig:Avalanche}, this region also contains the primary scintillation light; this contribution to the overall electron signal is constant throughout the experiment and is small with respect to the electron avalanche on the bottom PMT signal, of~($\mathcal{O}$(0.002~\%)). The primary scintillation process is therefore not considered throughout this analysis. The negative ion signal region extends from channel 900 to the end of the waveform. Signals within these regions below the charge threshold of -0.37~mV are integrated, including the two channels on either side of the threshold crossing location, to determine the signal intensity within each of these regions. An example of the integrated signal intensities for events with 54~Td and 45~Torr CF$_4$ are shown in Figure~\ref{fig:indSI}. These signal intensity distributions were fitted with Gaussian distributions and the mean of both the electron and negative ion distributions show a significant dependence upon the reduced field. The resulting fitted parameters of the integrated signal intensities for the range of reduced fields and pressures studied are presented in Figure~\ref{fig:DEA}.

The trends observed in the figure with respect to the reduced field show an increase in the electron avalanche signal from low fields to a maximum between 15-40~Td across all pressures. Following this maximum value, the intensity of the electron avalanche decreases, and between 20 and 35~Torr can become entirely absent. There is less electron avalanche suppression at higher pressures, indicating a possible pressure-dependent process that will be discussed in Section~\ref{Sec:Disc}.

The signal within the negative ion region is absent at reduced fields below 20~Td across all datasets, consistent with expectations of DEA from literature~\cite{Kurihara2000} and our Garfield++ simulations. With the exception of 15~Torr, as the reduced field increased beyond this 20~Td value, a proportional increase in the delayed signal is observed until a plateau is reached beyond the 70~Td point. The trends in the delayed signal are anti-correlated with the electron avalanche signal, indicative of proportional transfer between one signal region and the other. In some datasets, such as 30~Torr, the electron avalanche is completely diminished and the observable delayed signal exhibits approximately 70$\%$ of the peak electron intensity. 

%\begin{center}
\begin{figure*}[!htbp]
\centering
\includegraphics[width=\linewidth]{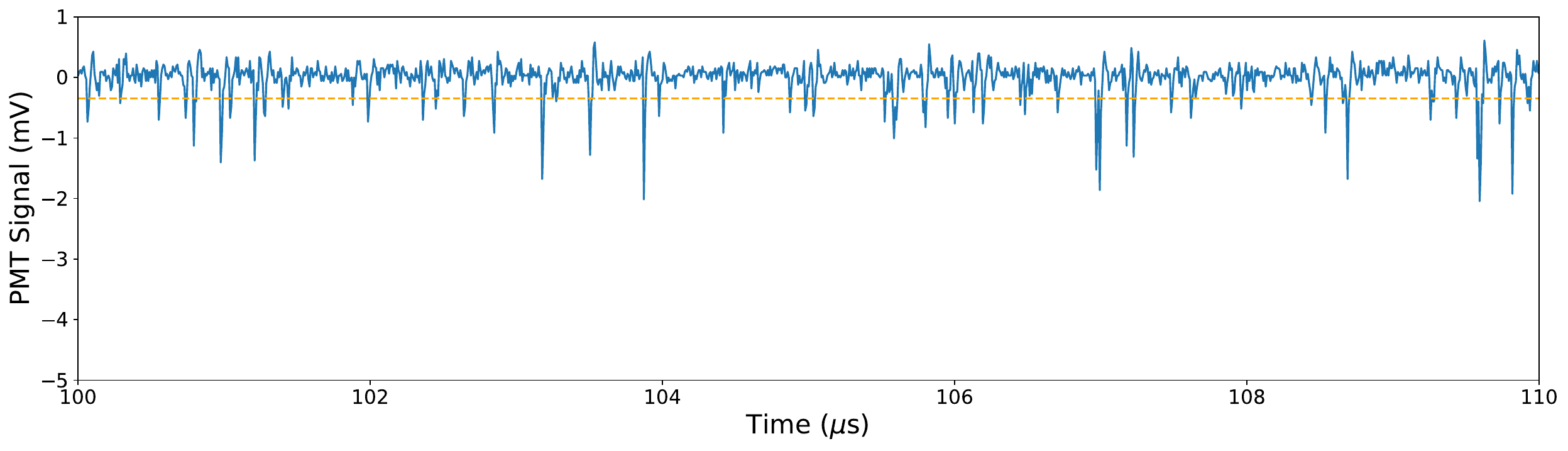}
\caption{A section of the delayed region of the PMT signal, taken under conditions above the DEA threshold, illustrates time resolvable individual peaks which we attribute to single negative ion avalanches. The yellow line shows the threshold used to analyse the arrival times and intensity distribution of these peaks.}
\label{fig:Singles}
\end{figure*}
%\end{center}

The lowest pressure observation of 15~Torr follows a similar trend to the 20-35~Torr data at low reduced fields for electron and delayed signals. However, above 150~Td both the electron and delayed signals continue to increase. This observation suggests that the field is sufficient to initiate impact ionisation in the drift field for these measurements, consistent with the predictions of the previous low pressure models.

\subsection{Time distribution}

\begin{figure}[!h]
\centering
\includegraphics[width=\columnwidth]{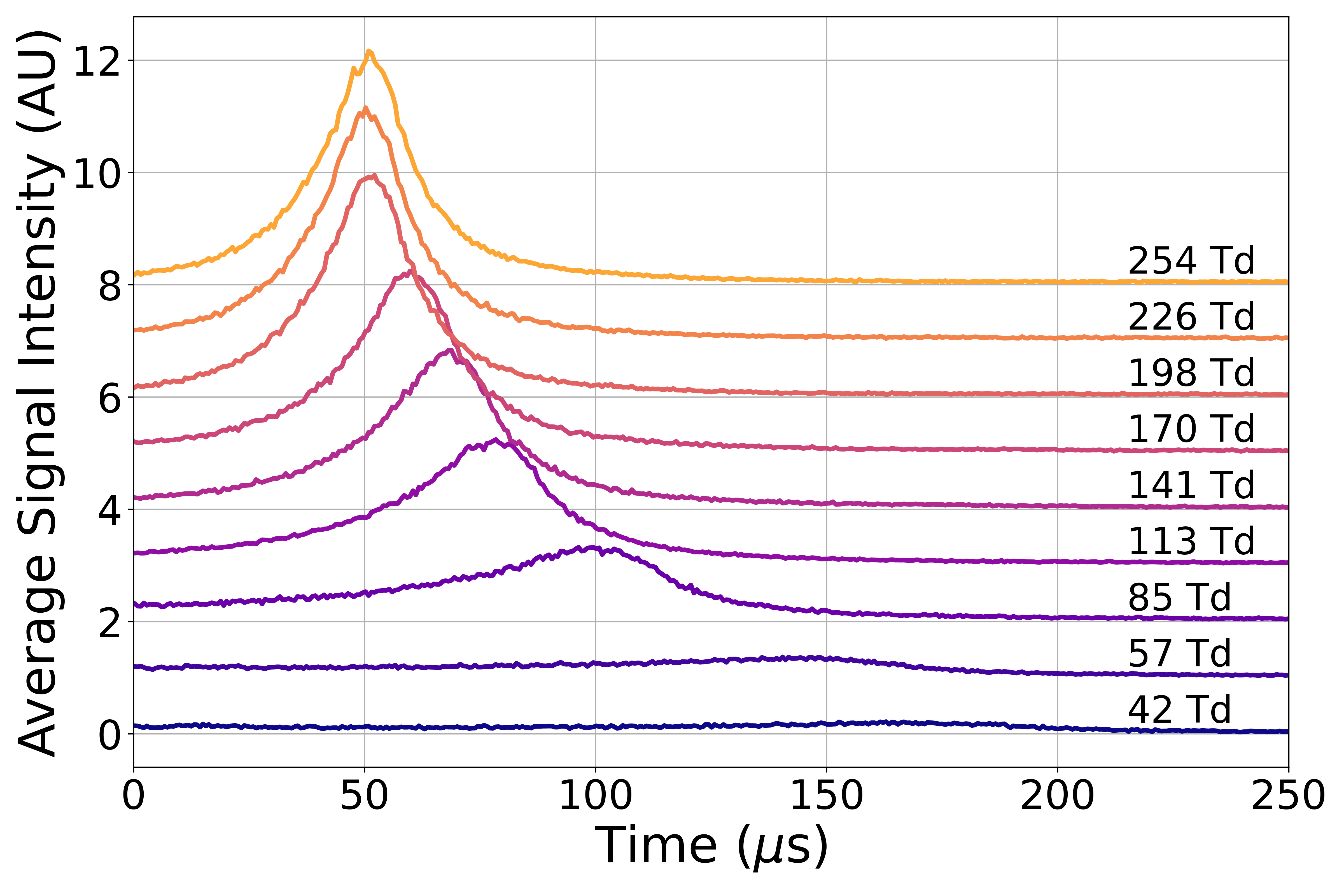}
\caption[]{Arrival time distributions above the DEA threshold, for various reduced fields as indicated in 20~Torr. The traces have been artificially offset for clarity. }
\label{fig:joydivision}
\end{figure}

The negative ion signal consists of individual peaks that are separable in time, owing to the fast time resolution of the PMT, as illustrated in Figure~\ref{fig:Singles}. The arrival time within the waveform of these peaks was extracted using the crossing locations for a threshold of -0.37~mV. The timing information for all traces in a given experimental configuration was combined to produce an averaged arrival time distribution. Figure~\ref{fig:joydivision} shows the average arrival time distributions for a range of reduced fields at a pressure of 20~Torr and similar sets of data can be made of the other observed pressures. In all cases, these resulting arrival time distributions are qualitatively similar to those observed by pulsed drift-tube experiments in previous studies of weakly attaching gas mixtures performed at $<$10~Torr~\cite{Dutton1985, JDutton1987a} including DEA to CF$_4$~\cite{JDutton1987,Dutton1987}. As predicted by simulation, the shape and location of the arrival time distribution varies as a function of the applied drift field, where an increased field shifts the peak arrival to earlier times. Arrival time distributions therefore encode information on the attachment coefficient and mobility of the negative ion species produced. 

\begin{figure}[]%% placement specifier
\centering%% For centre alignment of image.
\includegraphics[width=\columnwidth]{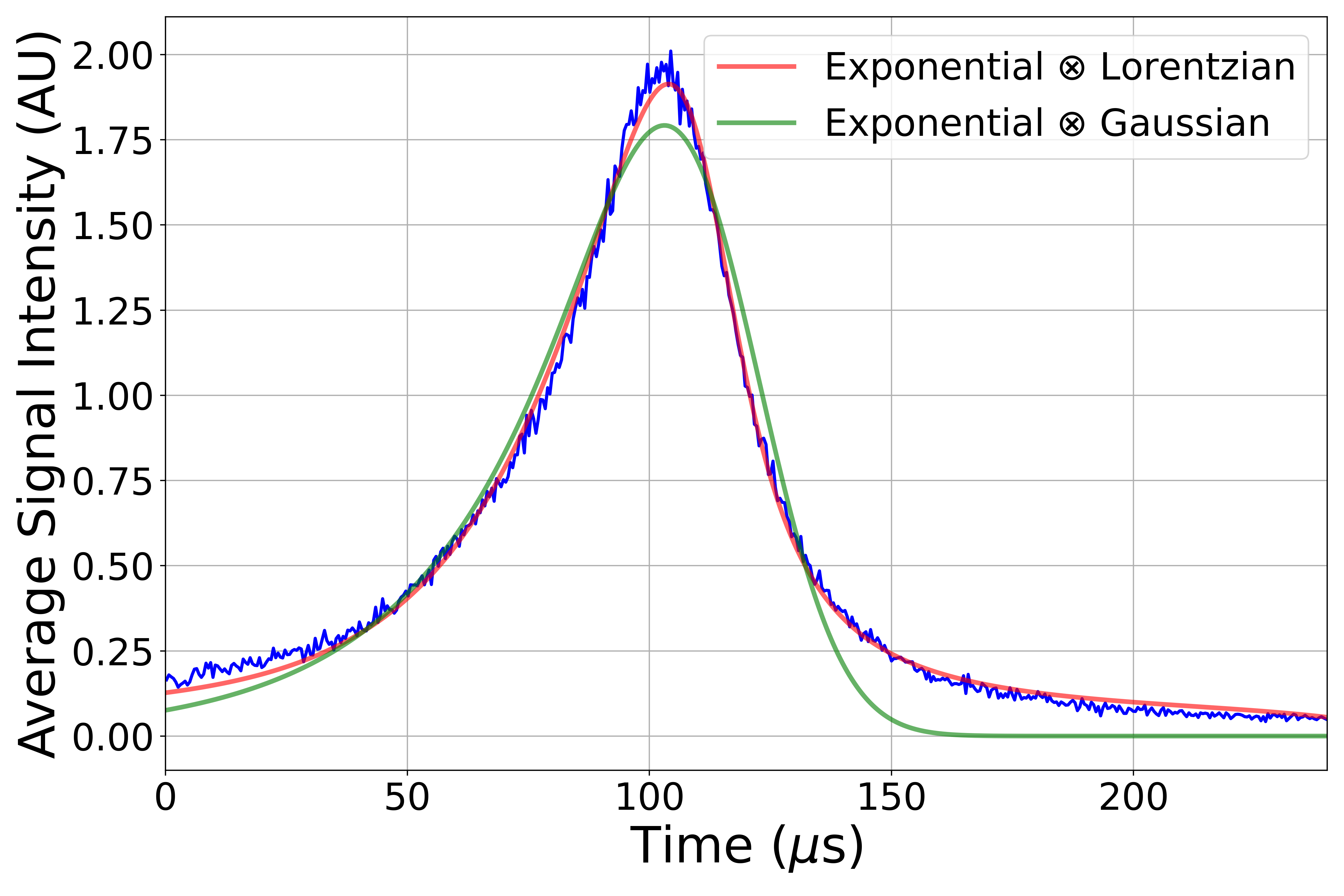}
\caption{The arrival time distribution with experimental data~(blue), fitted with an exponential convolved Lorentzian~(red) and Gaussian~(green) distributions. The exponential convolved Lorentzian fit is superior, consistent with simulation predictions (see text for details).}\label{fig:timedist}
\end{figure}

To build sufficient statistics, the arrival time distributions presented within this work are an accumulation of all the observed traces in a given measurement. This corresponds to a distribution of all possible track inclinations capable of triggering the 1~cm$^2$ photodiode. A Geant4~\cite{Agostinelli2003} simulation of the detector geometry was therefore used to model track inclinations and ionisation events from the collimated source which satisfies this condition. A distribution of z-positions~(drift direction) for primary ionisation locations was extracted for comparison with experimental timing information. This distribution was well-represented by a Lorentzian. In ideal conditions, where electron attachment immediately follows primary ionisation, the arrival time distribution should thus resemble a Lorentzian distribution. In situations where attachment is not immediate, the increased $v_d$ of electrons relative to ions means the unattached electrons traverse the gas in the drift direction prior to attachment, and thus present a tail skewing the arrival time distribution towards earlier times. This stochastic process is governed by the attachment cross section and follows an exponential distribution. Thus, a model of the arrival time distribution was chosen that consisted of an exponentially convolved Lorentzian distribution to account for these independent processes~(see Figure~\ref{fig:timedist}). 

The ion drift velocity was calculated from the mean of the fitted Lorentzian, making use of the known average drift distance of 1.8$\pm$0.1~cm. From this value, a field and pressure independent measure of the drift velocity, the reduced ion mobility~($K_0$), was determined~\cite{Marques2022}; 

\begin{equation}
    K_0 = \frac{v_d}{E\niss{drift}} \frac{N}{N_0}.
\end{equation}

Reduced mobilities measured in this work were in the range of 0.6-1.2~cm$^2$V$^{-1}$cm$^{-1}$ across the pressures studied. These values agree with the limited range of literature values for CF$_3^-$ and F$^-$ ion mobilities in CF$_4$ gas mixtures~\cite{JDutton1987}. This arrival time peak has been nominally associated with F$^-$ in previous work~\cite{JDutton1987}. However, comparison was difficult as the published values of 0.99-1.15 and 1.46-1.71~cm$^2$V$^{-1}$cm$^{-1}$ are not mass-identified and ambiguity exists over their exact chemical nature~\cite{JDutton1987}. Furthermore, previous publications have cautioned against the practice of using ion mobilities to identify signal carriers due to the possible influence of contamination, cluster formation, or other competing processes~\cite{JDutton1987, McDaniel1973}. 

\begin{figure}[]%% placement specifier
\centering%% For centre alignment of image.
\includegraphics[width=\columnwidth]{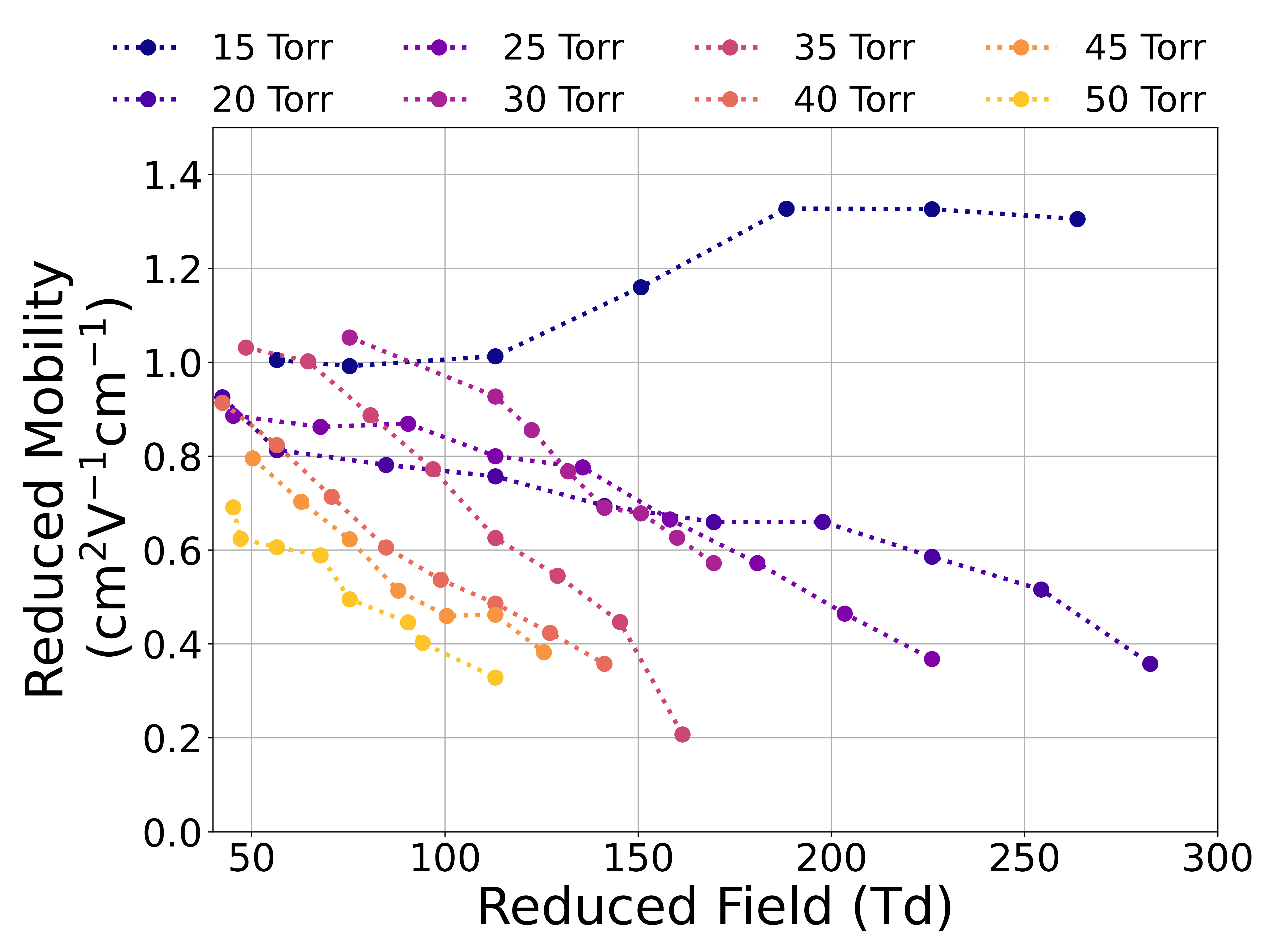}
\caption{Reduced mobilities of negative ions measured in pure CF$_4$. See text for details.}\label{fig:Mobility}
\end{figure}

\begin{figure}[]%% placement specifier
\centering%% For centre alignment of image.
\includegraphics[width=0.5\textwidth]{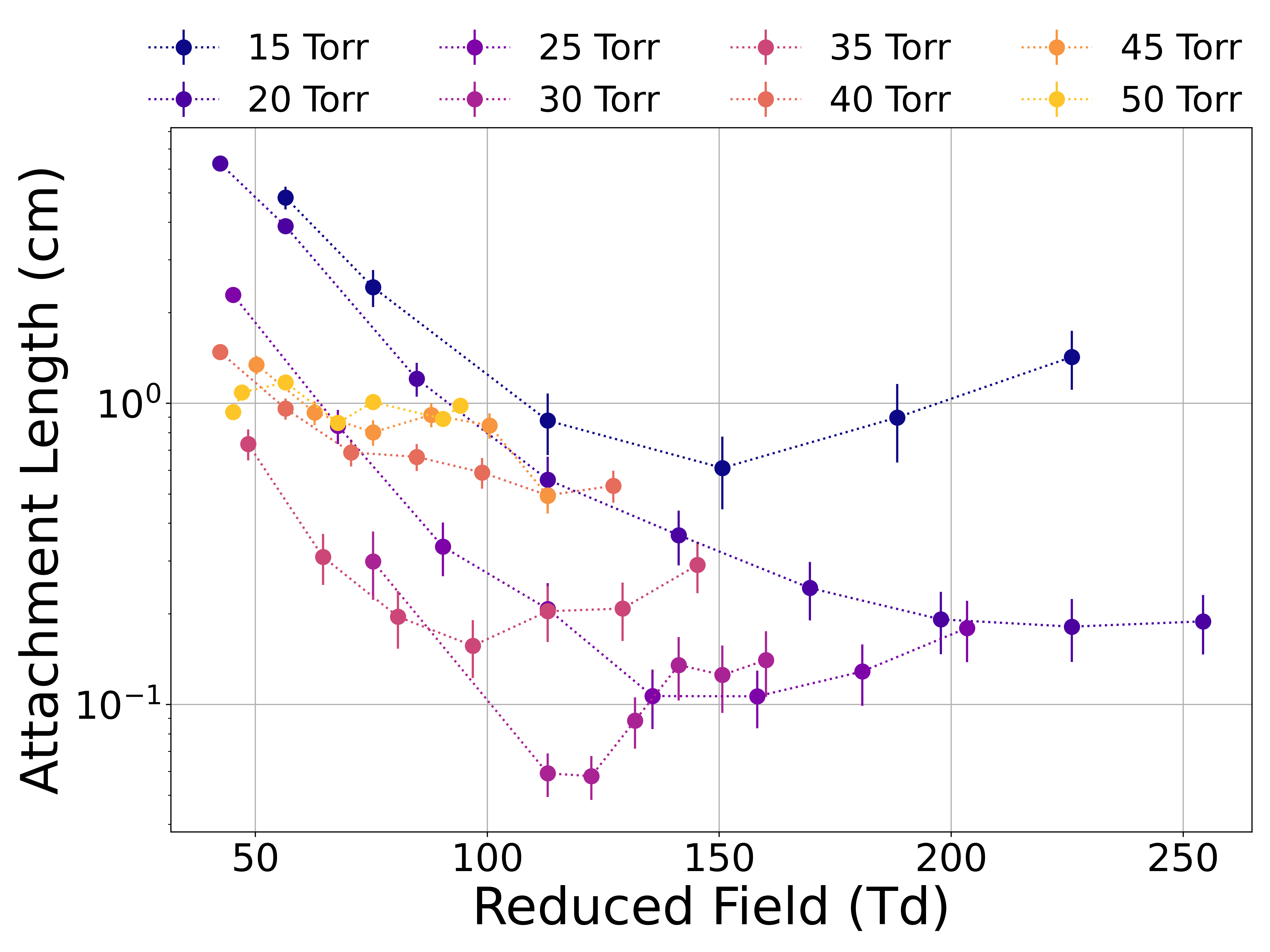}
\caption{Mean attachment lengths for the DEA negative ion production, extracted from the fits to the arrival time distribution. See text for details.}\label{fig:attachments}
\end{figure}

The rate of attachment within the gas volume is the attachment coefficient~($\eta$), from the fitted data, this is the exponential decay coefficient. Using the calculated ion mobility, a mean attachment length $\bar{\eta}$, is determined as the inverse of the exponential fit parameter expressed as cm$^{-1}$ from time using the drift velocity. The $\bar{\eta}$ values are shown in Figure~\ref{fig:attachments}. In the pressure range of 20-35~Torr, the $\bar{\eta}$ value decreases with increasing reduced field to a minimum, before increasing again. The location of this minimum appears to depend upon the gas pressure, with minima arriving at lower reduced fields for higher detector pressures. The smallest fitted attachment lengths for these pressures suggests attachment lengths below 1~mm are possible. Pressures above 35~Torr yield longer values of $\bar{\eta}$, increasing with greater pressures. This observation is in disagreement with the predictions from our Garfield++ simulation presented in Section~\ref{Sec:Simulation}, with potential causes of this trend discussed in Section~\ref{Sec:Disc}. 

% \begin{equation}
%    \eta = K \times t (check)
%    \label{Eq:eta}
% \end{equation}

\subsection{Single Carrier Gain Distribution}
\label{sec:singleions}
\begin{figure}[]%% placement specifier
\centering%% For centre alignment of image.
\includegraphics[width=\columnwidth]{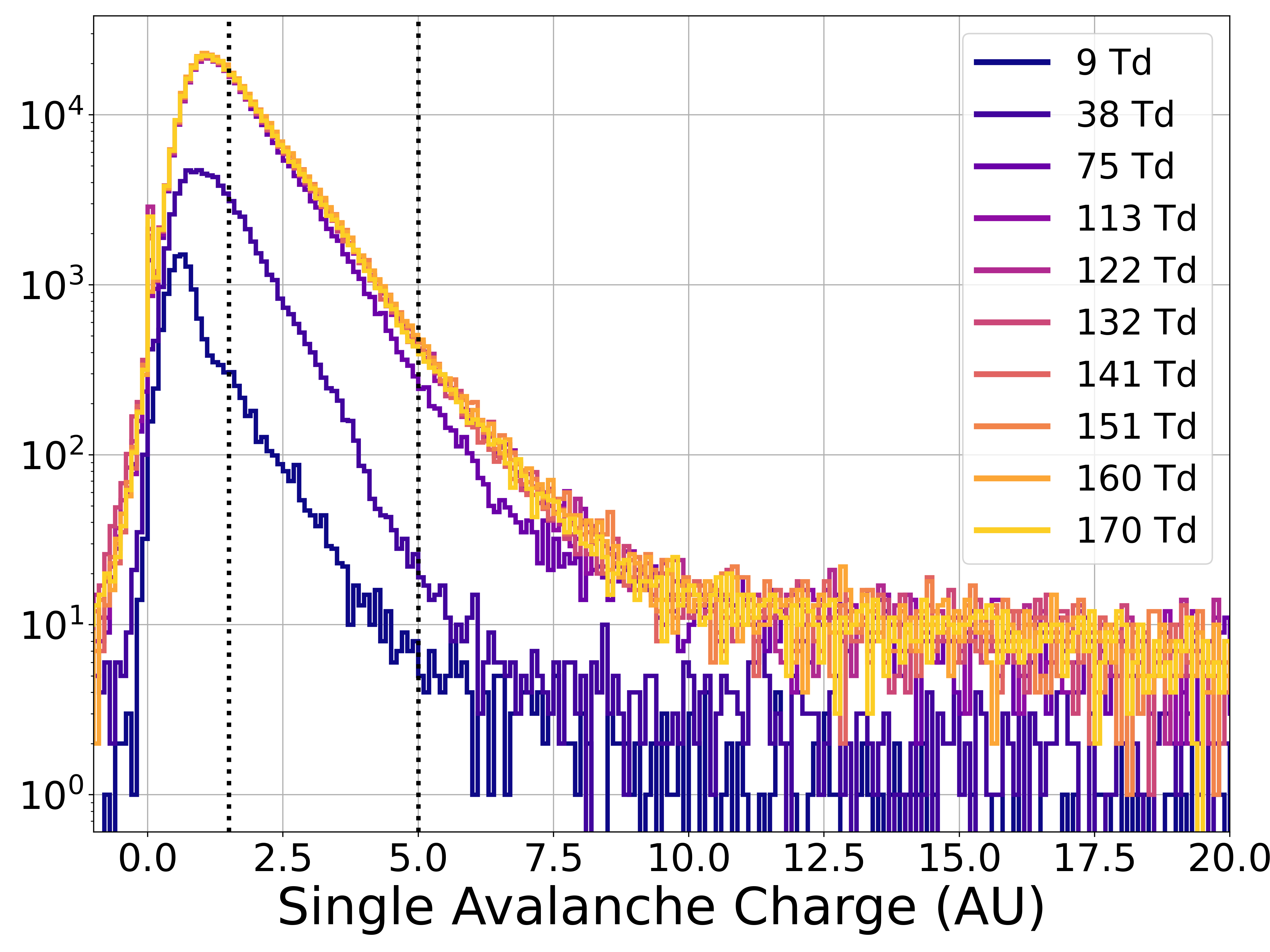}
\caption{Distribution of PMT signal charge observed from single negative ion avalanches, for 30~Torr CF$_4$ measurements. At high reduced fields, the distribution remains constant, and is well-represented by an exponential relationship.}\label{fig:SCG}
\end{figure}

The observation of isolated peaks in the negative ion signal region allows a measurement of the gain variation in the negative ion avalanche process. For individual peaks in the negative ion region, the charge of each peak was determined as the integrated signal exceeding the -0.37~mV threshold, with the immediate waveform samples on each side of this region included to account for the signal rise time. The distribution of observed waveform charge is presented in Figure~\ref{fig:SCG} for the 30~Torr measurements. At high reduced fields, in the region between the dashed lines of the figure, the distribution is consistent with an exponential Yule-Furry relationship~\cite{Blum2010}. Threshold effects are evident below this region and at higher charges single ion pileup is evident. Previous work investigating the single carrier gain curves in charge based detectors using O$_2^-$ ions observed a trend at low deposited charge deviating from this exponential relationship, which was attributed to the late detachment and consequential under-amplification of the ion~\cite{Sorensen2012}. This work shows no evidence of such structure, with the exception of fields below the DEA threshold~(see 9~Td curve in Figure~\ref{fig:SCG}). It is possible that the current work had insufficient sensitivity to the under-amplified component of the negative ion avalanche and could be further improved with an optimised gain structure. The avalanche process however, after surpassing the DEA threshold, is remarkably consistent between reduced fields.

\subsection{Cluster counting}

As discussed above, the fast PMT signal allows the observation of single ion avalanches at the readout plane. This configuration allows the detector to operate in a time expansion chamber regime~\cite{Walenta1979}, where counting the individual ion avalanches directly is expected to improve the energy resolution of the detector with respect to that of the conventional electron avalanche signal. This process is expected to be sensitive to the statistical fluctuation of the number of primary carriers presented by the Fano factor of the gas~($F$) and the probability of detecting a given carrier~($p$).

\begin{figure}[]%% placement specifier
\centering%% For centre alignment of image.
\includegraphics[width=\columnwidth]{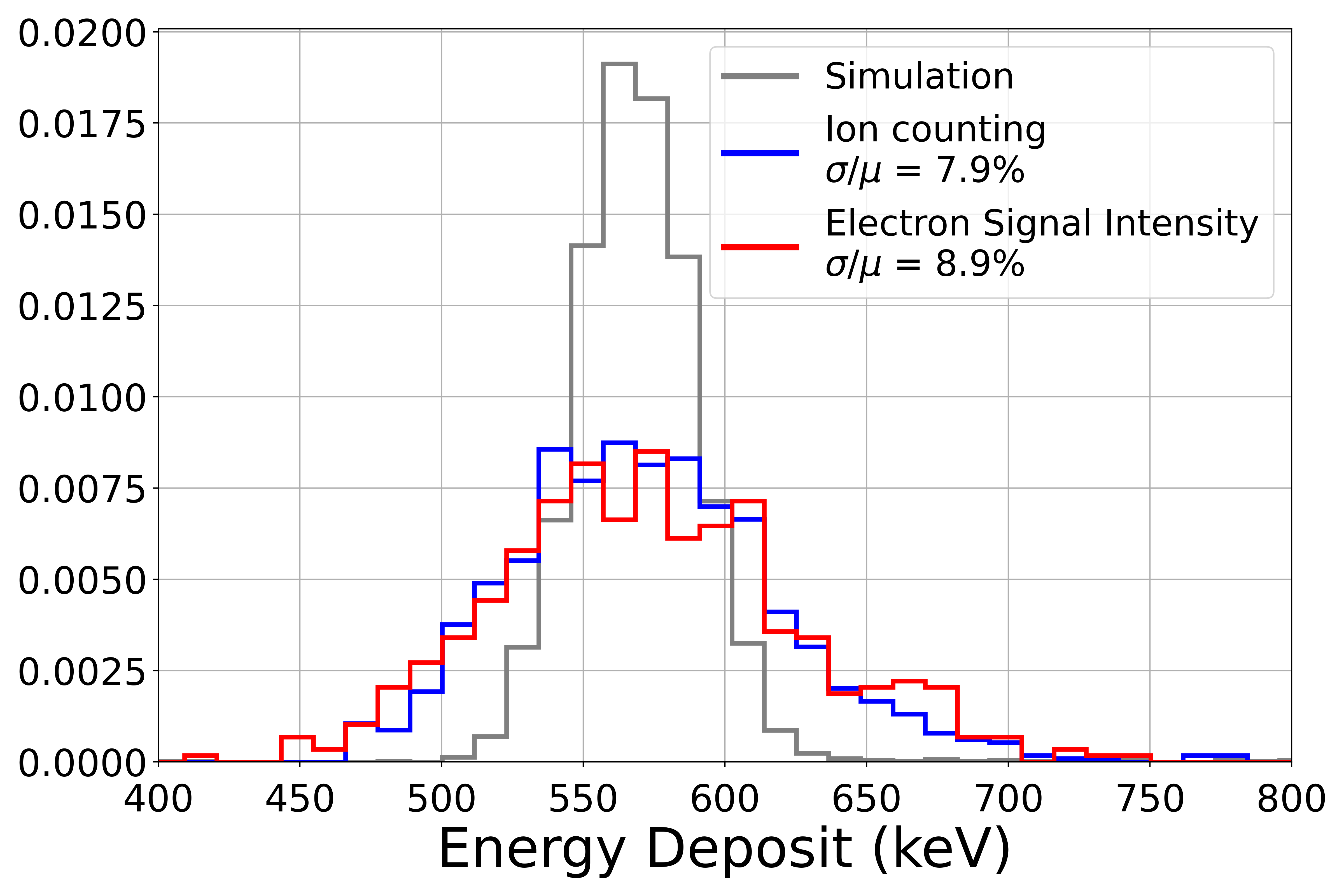}
\caption{Deposited energy distribution at 30~Torr. The simulated energy deposition~(grey) represents the best achievable resolution in a perfect detector. The single ion counting distribution~(blue) shows improved energy resolution when compared to the electron avalanche intensity distribution~(red).}
\label{fig:enres}
\end{figure}

Due to the varied operating pressures and incomplete energy deposition of the 5.5~MeV $^{241}$Am alpha particle between experiments, experimental data was compared to the simulated energy deposition for a given experimental setup using Geant4. As shown in Figure~\ref{fig:enres}, in the 30~Torr measurement where the best ion gain was present, an energy resolution of $\sigma/\mu=$7.9\% was observed compared to the $\sigma/\mu=$8.9\% value for the electron avalanche signal. This is a promising demonstration of the possibility of improved energy resolution when operating in this mode, and more studies are needed to optimise the effect at higher avalanche gain.

%##############################################################################################
\section{Discussion}
\label{Sec:Disc}

Known detector effects of PMT after-pulsing~\cite{Torre1983, Akchurin2007} and ion feedback~-\cite{Gasik2024} were considered as possible alternate explanations of the negative ion signal. PMT after-pulsing can be excluded as the negative ion signals arrive over 100s of $\mu$s, and have a dependence on E$\niss{drift}$ whereas typical PMT after-pulsing occurs within a few $\mu$s and is independent of the detector conditions~\cite{Torre1983, Akchurin2007}. Ion feedback~\cite{Gasik2024} produces delayed signals following an avalanche, and could plausibly occur on the timescales of our delayed signals via the liberation of additional electrons at the cathode by the positive ions produced in the initial avalanche. However, the absence of any delayed signal at low drift fields out to 2~ms~(corresponding to a limit on reduced mobilities of $<$0.35~cm$^2$V$^{-1}$s$^{-1}$) is inconsistent with an ion feedback hypothesis for F$^+$ and CF$_3^+$, which have ion mobilities of 1.55 and 0.92~cm$^2$V$^{-1}$s$^{-1}$, respectively~\cite{Stojanovic2014}. Furthermore, ion feedback requires an initial electron avalanche to produce positive ions, and our delayed signals are anti-correlated with the presence of the primary electron avalanche, with strong delayed signals measured in the absence of an electron avalanche. The consilience of evidence suggests that the delayed signals are the result of DEA interactions with CF$_4$ molecules.

\subsection{Gain}
% Gain discussion
Previous works investigating NID in low pressure CF$_4$ have found thin GEMs too unstable to operate due to the high gains~(10$^4$) required to observe signals, and thus have preferred the use of thick GEMs~\cite{Phana}. In this work, despite using thin GEMs, modest gains were achieved for the electron avalanche and a negative ion signal was observed at comparable overall gain~(70~$\%$ of the electron signal intensity) from the same amplification field~($\Delta V\niss{GEM}$). This contrasts with previous reports of the gain relative to electrons measured for negative ions in SF$_6$~\cite{McLean2024_} under the same amplification fields, although a direct comparison is difficult due to the electronegative nature of SF$_6$-based NID gas mixtures preventing simultaneous electron-NID comparisons. The lower NID gain in SF$_6$ may therefore be symptomatic of parasitic effects that lead to a reduced Townsend coefficient, imparted by the electronegative fill gas during the avalanche. Alternatively, the relatively higher gain observed in this study may be characteristic of more favourable electron detachment from the negative ions. This possible beneficial effect is surprising as both expected ions, F$^-$ and CF$_3^-$, have higher electron affinities~(3.45~eV and 1.8~eV, respectively~\cite{Macneil1970}) than previously observed negative ions such as SF$_6$ and CS$_2$~(1.06 and 0.5~eV, respectively~\cite{Christophorou1998, Lifshitz1973}). The disparity between these two observations is an area for future investigation.

\subsection{Pressure Dependence}

As outlined in Section~\ref{Sec:Simulation}, the DEA process is a two-body interaction and was expected to have shorter attachment lengths with increased pressure. Our measurements concur with this expectation until above 35~Torr, where transfer from electrons to ions was incomplete and fits suggested a longer attachment length with increasing pressure for a given reduced field. A possible explanation for this result arises from collisional autodetachment. Electron attachment interactions with CF$_4$ must first traverse a temporary excited state CF$_4^{-*}$, formed by;   

\begin{align}
    \mathrm{CF}_4 + e^- &\leftrightarrow \mathrm{CF}_4^{-*}.
    \label{eq:belowDEA}
\end{align}

As the resulting ion is unstable, it will not stabilise through collisions and must undergo auto-detachment unless sufficient energy is available to dissociate into stable ion fragments from the unstable parent species as shown in Equations~\ref{eg:cf3} and~\ref{eg:f}. However, at sufficiently high pressures, the mean time between molecular collisions may be shorter than the CF$_4^{-*}$ dissociative lifetime, allowing interactions with neutral CF$_4$ such as;  
\begin{align}
    \mathrm{CF}_4^{-*} + \mathrm{CF}_4 &\rightarrow \mathrm{CF}_4^{-*} + \mathrm{CF}_4^*.
    \label{eq:collisionalenergytransfer}
\end{align}

Such inelastic energy transfers to neutral CF$_4$ are likely to proceed readily, due to its many low lying vibrational and rotational states~\cite{Christophorou1996}. This interaction could plausibly reduce the CF$_4^{-*}$ ion's excitation below the DEA threshold, leading to autodetachment. Alternatively, the negative EA of the CF$_4$ molecule implies a collisional interaction may have sufficient energy to immediately detach the electron such as;

\begin{align}
    \mathrm{CF}_4^{-*} + \mathrm{CF}_4 &\rightarrow 2\mathrm{CF}_4^* + e^-.
    \label{eq:collisionaldetachment}
\end{align}

Both hypothesised interactions become an effective three-body process, so a $P^2$ dependence would be expected. This could potentially limit the operational pressures of DEA-based detectors with pure CF$_4$ and will be investigated in future work. 

\section{Conclusion and future work}

Electronegative gases have been added to gas based detectors to improve spatial resolution through negative ion drift and increase stability by reducing feedback mechanisms. However, the conventional approach of adding electronegative gases to detectors presents undesired detector responses or environmental impacts. Using dissociative electron attachment as a mechanism to produce negative ion charge carriers from the fragments of the fill gas, negative ion drift has been demonstrated within this work using a conventional drift gas, CF$_4$, in an optical time projection chamber, for the first time. The negative ion formation is initiated by increasing the drift field above 20~Td, opening the possibility for applications where negative ion and electron carrier drift modes can be chosen interchangeably within a single detector. DEA attachment lengths of less than 1~mm were observed. The measurements are consistent with the DEA threshold expected from simulation, as well as simulated predictions of a shortening attachment length with increasing pressure below 35 Torr. Above this pressure, measurements determined longer attachment lengths, indicating the possible presence of processes that compete with DEA. The negative ion gains were as high as 70\% of the electron gain, which is anomalously high compared to other negative ion studies. Finally, the excellent timing properties of the photomultipliers allowed avalanches from individual negative ions to be measured, opening a probe to the single carrier gain curve and a demonstration of a potential improvement to the energy resolution by counting negative ion avalanches, relative to the electron avalanche signal intensity. This work presents a method for future gas-based detectors to both explore the attachment and detachment processes in gasses and provide an avenue for improved spatial and energy resolution in time projection and time expansion chambers. 

\section{CRediT authorship contribution statement}

\textbf{L.J. McKie:} Conceptualization, Methodology, Software, Resources, Investigation, Formal Analysis, Writing -- Original draft. \textbf{L.J. Bignell:} Methodology, Formal Analysis, Visualisation, Writing -- Review and editing, Supervision, Funding Acquisition. \textbf{N. Adams:} Software, Investigation. \textbf{V.U. Bashu:} Investigation. \textbf{F. Dastgiri:} Resources, Investigation. \textbf{G.J. Lane:} Writing -- Review and editing, Supervision, Funding Acquisition.

\section{Acknowledgements}
The authors offer their sincere thanks to S. Buckman and J. Machacek for valuable conversations and feedback on the manuscript. The authors wish to acknowledge the technical staff of the ANU Department of Nuclear Physics and Accelerator Applications and the Heavy Ion Accelerator Facility for their assistance in this research.

%\section*{References}

%\def\refname{\vadjust{\vspace*{-1em}}} %Please don't do this in a real paper.
\bibliographystyle{IEEEtran}
\bibliography{121224_3}

\end{document}